%% file: review_EPJC.tex
\documentclass[pdftex,smallcondensed,epjc3]{svjour3}          

\usepackage[T1]{fontenc}
\smartqed  

\usepackage{subfig}
\usepackage{graphicx}
\usepackage{mathptmx}      
\usepackage{flushend}
\usepackage{color}
\usepackage[numbers,sort&compress]{natbib}
\usepackage[colorlinks,citecolor=blue,urlcolor=blue,linkcolor=blue]{hyperref}
\usepackage{multicol}
\usepackage{bbold}
\usepackage{amsmath}
\usepackage{amssymb}
\usepackage{xspace} 
\usepackage{slashed}
\usepackage{units}
\usepackage{eufrak}
\usepackage{bbm}

\journalname{Eur. Phys. J. C}

\newcommand{\MeV}{{\rm \,MeV}}
\newcommand{\GeV}{{\rm \,GeV}}

\newcommand{\cm}{{\rm \,cm}}

 \def\be   {\begin{equation}}   \def\ee   {\end{equation}}
 \def\bea  {\begin{eqnarray}}   \def\eea  {\end{eqnarray}}
 \def\nn{\nonumber}
 
 \def\x {\chi}
 \def\xbar {\bar\chi}
 \def\qbar{\bar q}
\def\gmd{\gamma_\mu}
\def\gmu{\gamma^\mu}
\def\g5{\gamma_5}
\newcommand{\Op}[2][]{ \ensuremath{\mathcal{O}^{\rm #1}_{\rm #2} }}

\newcommand{\Mstar}{\ensuremath{M_{*}}\xspace}
\newcommand{\mMed}{\ensuremath{M_{\rm{med}}}\xspace}
\newcommand{\Zprime}{\ensuremath{Z^\prime}\xspace}
\newcommand{\gDM}{\ensuremath{g_{\chiDM}}\xspace}
\newcommand{\chiDM}{\ensuremath{\chi}\xspace}
\newcommand{\gq}{\ensuremath{g_{\rm q}}\xspace}
\newcommand{\mDM}{\ensuremath{m_{\chiDM}}\xspace}
\newcommand{\Mcut}{\ensuremath{M_{\text{cut}}}\xspace}
\newcommand{\gstar}{\ensuremath{g_{*}}\xspace}
\newcommand{\Qtr}{\ensuremath{Q_{\rm tr}}\xspace}
\newcommand{\tev}{\ensuremath{\mathrm{\,Te\kern -0.1em V}}\xspace}

\newcommand{\sigSI}{\ensuremath{\sigma_{\rm SI}}\xspace}
\newcommand{\sigSD}{\ensuremath{\sigma_{\rm SD}}\xspace}
\newcommand{\sigv}{\ensuremath{\langle\sigma v\rangle}\xspace}
\newcommand{\cgoth}{\ensuremath{\mathfrak{c}}\xspace}

\begin{document}

\title{Simplified Models vs. Effective Field Theory Approaches in Dark Matter Searches}

\author{Andrea De Simone 
\thanksref{addr1} 
\and
Thomas Jacques 
\thanksref{addr1} 
}

\institute{SISSA and INFN Sezione di Trieste, via Bonomea 265, I-34136 Trieste, Italy\\
\email{\texttt{\href{mailto:andrea.desimone@sissa.it}{andrea.desimone@sissa.it}, \href{mailto:thomas.jacques@sissa.it}{thomas.jacques@sissa.it}}}
\label{addr1}
}

\date{13 July 2016}

\maketitle

\begin{abstract}
In this review we discuss and compare the usage of simplified models and 
Effective Field Theory  (EFT) approaches in dark matter searches.
We provide a state of the art description on the subject of EFTs and simplified models, especially in the context of collider searches for dark matter, but also with implications for direct and indirect detection searches, with the aim of constituting a common language for future comparisons between different strategies. 
The material is presented in a form that is as self-contained as possible, so that it may serve as an introductory review for the newcomer as well as a reference guide for the practitioner.
\end{abstract}

\newpage

\linespread{1.4}
\tableofcontents


\input{sections/intro.tex}

\input{sections/EFTs.tex}

\input{sections/simp.tex}

\input{sections/conclusion.tex}


\begin{acknowledgements}
We are grateful to our collaborators  Giorgio Busoni, Johanna Gramling, Enrico Morgante and Antonio Riotto,  with whom we co-authored papers on this subject.
We also thank the members of the LHC Dark Matter Working Group for very interesting discussions.
\end{acknowledgements}


\bibliographystyle{JHEP}

\bibliography{bibliography}

\end{document}

%% file: sections/intro.tex
\section{Introduction}

The existence of a Dark Matter (DM) component of the universe is now firmly established, receiving observational support from gravitational effects both on astrophysical scales and on cosmological scales.
The DM abundance is precisely known and 
 can be expressed in terms of the critical energy density as $\Omega_{\rm DM}h^2=0.1196\pm 0.0031$
\cite{1303.5076}, which corresponds to about one quarter of the total energy  content of our universe. 
Besides this, almost no other experimental information is available about  the nature of Dark Matter and its interactions with the Standard Model (SM) of particle physics.

The paradigm for the DM particle which has been most thoroughly studied, especially motivated by the attempts to solve the hierarchy problem such as Supersymmetry,  is that
of a Weakly Interacting Massive Particles (WIMP), with weak-scale interactions and masses in the range of about GeV-TeV. 
In this review we will stick to the WIMP paradigm, so we will use DM and WIMP interchangeably.

Experimental searches for WIMPs 
attack the problem from very different angles, in an attempt to (directly or indirectly) probe the nature of the DM particle.
Broadly speaking, the search strategies currently ongoing proceed in
 three main directions:
(1) collider searches,  identifying the traces of direct production of DM in particle colliders; 
(2) direct searches,  looking for the scattering events of DM with heavy nuclei in a shielded underground laboratory;
(3) indirect searches,  detecting the final products  of DM annihilations in the galaxy or in the Sun, such as gamma-rays or neutrinos.

The benefit of exploiting the complementary interplay among these different approaches is to improve the discovery potential in a significant way.
As this interplay is gaining more and more importance in recent years, the need for a common language into which to translate the results of the different searches has become more pressing. 
The efforts to develop more model-independent approaches to DM searches (especially for collider physics) stimulated a vast literature on the subject
\cite{hep-ph/0011335,hep-ph/0405097,hep-ph/0403004,hep-ph/0503117,hep-ph/0605188,0803.4005,0808.0255,0808.3384,0809.2849,0811.0393,0909.0520,0912.4511,0912.4722, 1002.4137,1003.1912,1005.1286,1005.3797,1005.5651,1008.1591,1008.1783,1008.1796,1009.0008,1009.0618,1011.2310,1012.2022, 1102.3024,1103.0240,1103.3289,1104.1429,1104.5329,1106.3097,1107.2048,1107.2118,1108.0671,1108.1196,1108.1800, 1109.3516,1109.4398,1109.4872,1110.4405,1111.2359, 1111.2835,1111.4482,1112.3299,1112.5457, 1201.0506,1201.3402,1201.4814,1202.2894,1203.1662,1203.2064,1203.3542,1204.3839,1205.3169,1206.0640,1207.1431,1207.3971,1208.4361,1208.4605,1209.0231,1210.0195,1210.0525,1212.2221,1212.3352, 1301.1486,1302.3619,1303.3348,1303.6638,1306.4107,1307.1129,1307.2253,1307.5740,1307.6277,1307.8120,1308.0592,1308.0612,1308.2679,1308.6799,1309.3561,1309.4017,1310.4491,1310.6047,1311.1511,1311.5896,1311.6169,1311.7131,1312.0009,1312.2592, 1312.5281, 1401.0221,1402.1173,1402.1275,1403.6734,1404.0051,1407.7494, 1408.2745,1408.3583,1408.5046,1410.4031,1410.8812,1411.1559,1412.0520, 1501.00907,1502.01518,1503.07874,1504.01395,1504.03198,1506.07475,1507.08294,1509.01587,1511.07463,1603.05994}.

The approach of using Effective Field Theory (EFT) is based on describing the unknown DM interactions with the SM in a very economical way. This has attracted significant attention, especially because of its simplicity and flexibility which allows it to be used in vastly different search contexts. 
Unfortunately, the validity of this approach, as far as the collider searches for DM are concerned, has been questioned 
\cite{1111.2359,1112.5457,1203.4854,1307.2253,1308.6799,1402.1275,1405.3101} 
and the limitations to the use of EFTs are by now recognized by the theoretical and experimental communities 
\cite{1409.2893,1409.4075,1506.03116,1507.00966,1603.04156}.

Certainly,  one way out of this \textit{impasse} is to resort to full-fledged models of new physics, comprising a DM candidate. 
For example, models connected to the solution of the hierarchy problem,
such as supersymmetric models or models with a composite Higgs, are already being thoroughly studied.
These kinds of searches for DM within more complete frameworks of particle physics have been and are currently the subject of a great deal of research.
The results often play the role of benchmarks to be used among different communities
of DM hunters.
On the other hand, more fundamental frameworks necessarily involve many parameters.
Therefore, the inverse problem, i.e. using experimental results to understand the theory space, necessarily involves a large number of degeneracies. This is a particularly severe problem for DM, for which the only precisely known property is the relic abundance.

A ``third-way'' between these two extremes, the effective-operator approximation and complete
ultraviolet models, is possible and is indeed convenient.

The logic behind the so-called \emph{simplified models} \cite{hep-ph/0703088, 0810.3921, 1105.2838} is to expand the effective-operator interaction to include the degrees of freedom of a ``mediator'' particle, which connects the DM particle with the 
Standard Model sector. This amounts to assuming that  our ``magnifying glass'' (the LHC or a future collider) is powerful enough to be able to go beyond the coarse-grained picture provided by EFT and resolve more microscopic -- though not all --  details which were integrated out.
In the limit of sufficiently heavy mediators, the EFT situation is recovered.

This way of proceeding has appealing features as well as limitations. Of course, despite being simple and effective, this is not the only way to go. In fact one may look for alternative scenarios which, while not fully committing to specific models, still offer  diversified phenomenology, e.g. along the lines of the benchmarks in Ref.~\cite{1402.6287}.
Furthermore, the simplified model approach may look rather academic, as these models are unlikely to be a realistic fundamental theory.

On the other hand, simplified models retain some of the virtues of the other extreme approaches: a small number of manageable parameters for simpler search strategies, and close contact with ultraviolet completions, which reduce to the simplified models in some particular low-energy limit.
Moreover, one can exploit the direct searches for the mediator as a complementary tool to explore the dark sector.

In this review, we summarize the state of the art of DM searches using EFT and simplified models. Our focus will be primarily on collider searches but we will also discuss the connections with direct and indirect searches for DM.
In Section \ref{sec:eft} we highlight the virtues and drawbacks of the EFT approach, and provide the formulae which are necessary to establish the links among collider/direct/indirect searches, so that a unified picture emerges.
In Section \ref{sec:simp} we shift the attention to the simplified models.
A classification of these models according to the quantum numbers of the mediator and DM particles and the tree-level mediation channel,  is used as a guideline for the discussion of the different kinds of model.
We also propose an easy-to-remember nomenclature for the simplified models and point out which ones still need further investigation.

%% file: sections/EFTs.tex
\section{Effective Field Theories: virtues and drawbacks}
\label{sec:eft}

Given that the particle nature of DM and its interactions are still unknown, it is important
that analyses of experimental data include constraints that cover as broad a range of DM models
as possible in a way that is as model-independent as possible. Whilst the EFT approach does have limitations, it remains a powerful tool to achieve this goal.
This approach should be complemented by both  limits on the raw signal, and constraints on models which capture the full phenomenology of well-motivated UV-complete DM models, but none of these approaches should stand in isolation. 

The EFT approach involves reducing the interactions between DM and the SM fields down to contact interactions, described by a set of non-renormalisable operators, for example,
\be
\mathcal{L}_{\rm EFT} = \frac{1}{\Mstar^2}\left(\bar{q} q\right)\left(\bar{\chi}  \chi\right) \, .
\ee
In this case, a fermionic DM particle $\chi$ and SM quark $q$ are coupled via a scalar interaction. The strength of the interaction is governed by an energy scale $\Mstar$, taken to the appropriate power for this dimension-6 operator

\begin{figure}
	\centering
	\includegraphics[width=0.4\textwidth]{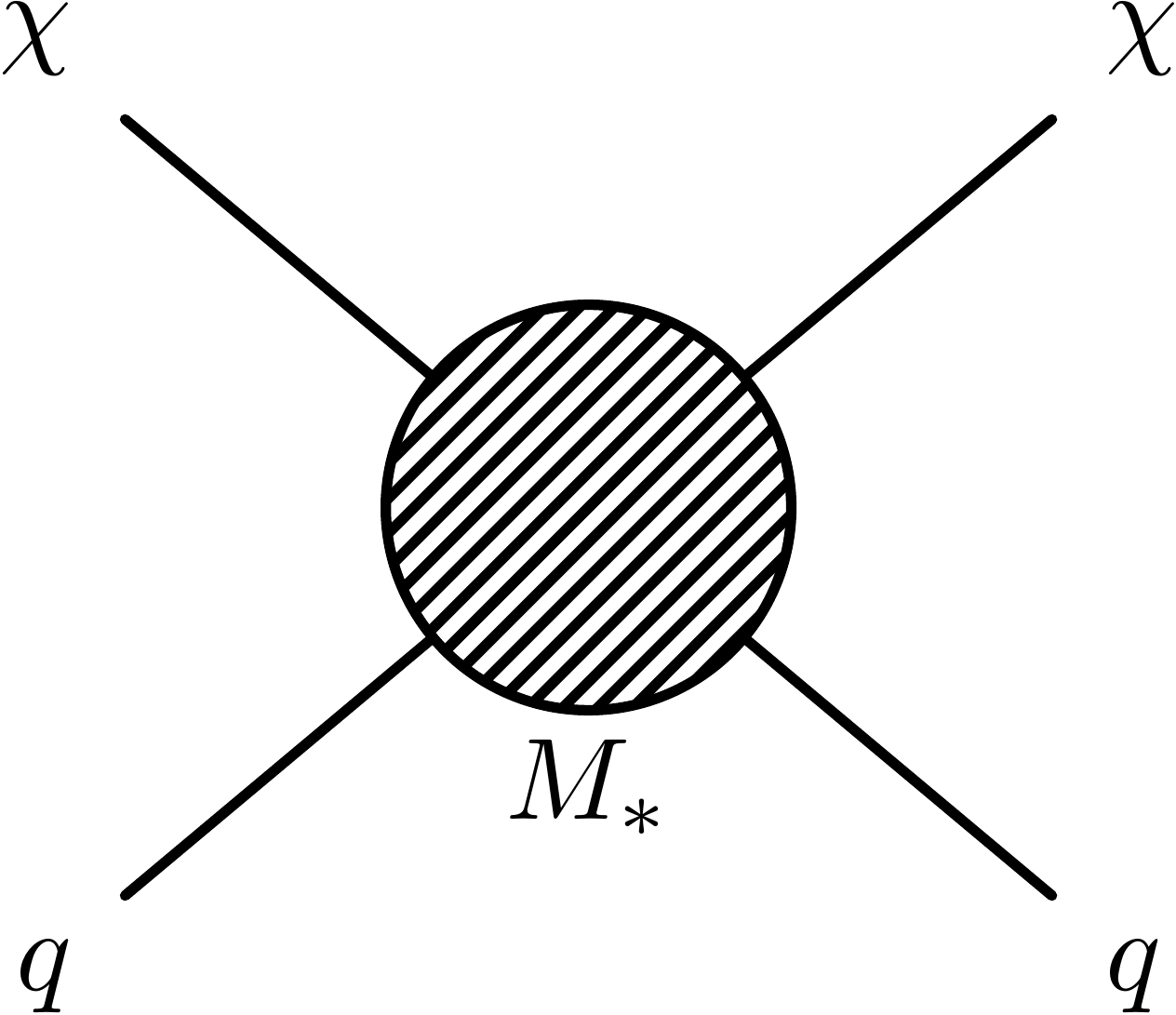}
	\caption{Schematic of an EFT interaction between DM and the SM}
	\label{fig:feynblob}
\end{figure}

The beauty of the EFT approach is that each operator and energy scale describe a range of processes, depending on the direction of the arrow of time in Fig.~\ref{fig:feynblob}: DM annihilation, scattering, and production can all described by the same operator. As we will describe in more detail in the following section, calculations using these operators correspond to taking an expansion in powers of the energy scale of the interaction, along the lines of $E^n/\Mstar^n$, and truncating. Therefore EFT calculations are a consistent description of a higher-order process if and only if the energy scale of the interaction is small compared to the energy scale $\Mstar$. Therefore the EFT description is strongest when there is a clear separation between the energy scales of the operator and the interaction.
 In  the context of DM searches, there are several situations where the EFT approach is absolutely
 solid. In indirect searches, for example, 
the energy scale for the non-relativistic annihilation of DM particles 
in the halo is  of the order of the DM mass $m_{\rm DM}$;
direct DM searches probe the non-relativistic DM-nucleon operator, where the energy transfer is of the order of MeV.
Therefore, as long as the mediator is heavier than $\mathcal{O}$(MeV) ($\mathcal{O}(m_{\rm DM})$), EFTs can provide a consistent description of (in)direct detection, as we outline in Sections~\ref{sec:eft-dd} and \ref{sec:eft-id}.

However, the situation is substantially different in LHC searches for DM.
In fact,  effective operators are a tool to describe
the effects of heavy particles (or `mediators') in the low energy theory where these particles
have been integrated out. 
But the LHC machine delivers scattering events at  energies so high, that
they may directly produce the mediator itself. Of course, in this case 
the EFT description fails. While EFT analyses remain a useful tool for LHC searches,  this simple point calls for a careful and consistent use of the EFT,
checking its range of validity, in the context
of DM searches at the LHC.

\subsection{Effective Field Theories for collider searches\label{sec:eft-lhc}}

EFTs are useful at colliders as a parameterisation of missing energy searches. If DM is produced alongside one or more energetic SM particles, then the vector sum of the visible transverse momentum will be non-zero, indicating the presence of particles invisible to the detector, such as neutrinos, DM, or long-lived undetected particles. 

The most relevant operators for collider searches are the relativistic DM-quark and DM-gluon operators, shown in Tables \ref{DList} and \ref{MList} for Dirac and Majorana fermionic DM, and Tables \ref{CList} and \ref{RList} for complex and real scalar DM respectively, where $\tilde{G}^{\mu\nu} \equiv \epsilon^{\mu\nu\rho\sigma}G_{\rho\sigma}$. 
The parameter \Mstar is of course independent for each operator, and in principle for each flavor of quark, although \Mstar is generally assumed to be flavor-universal in collider studies, in order to avoid issues with flavor constraints, such as flavor-changing neutral currents.

\begin{table*}
  \begin{tabular*}{\textwidth}{@{\extracolsep{\fill}}lrrc}
  \hline
Label & \multicolumn{1}{c}{Operator} & \multicolumn{1}{c}{Usual coefficient} & Dimension \\
 \\
\hline
\Op{D1} & $\xbar  \x \qbar q$ & $m_q/\Mstar^3$ & 6 \\
\Op{D2} & $\xbar i \g5 \x \qbar q$ & $m_q/\Mstar^3$ & 6 \\
\Op{D3} & $\xbar   \x   \qbar  i \g5  q$ & $m_q/\Mstar^3$ & 6 \\
\Op{D4} & $\xbar i \g5 \x \qbar i \g5 q$ & $m_q/\Mstar^3$ & 6 \\
\Op{D5} & $\xbar \gmu \x \qbar \gmd q$ & $1/\Mstar^2$ & 6 \\
\Op{D6} & $\xbar \gmu \g5 \x \qbar \gmd q$ & $1/\Mstar^2$ & 6 \\
\Op{D7} & $\xbar \gmu \x \qbar \gmd \g5 q$ & $1/\Mstar^2$ & 6 \\
\Op{D8} & $\xbar \gmu \g5 \x \qbar \gmd \g5 q$ & $1/\Mstar^2$ & 6 \\
\Op{D9} & $\xbar \sigma^{\mu \nu} \x \qbar \sigma_{\mu\nu} q$ & $1/\Mstar^2$ & 6 \\
\Op{D10} & $\xbar i \sigma^{\mu \nu} \g5 \x \qbar \sigma_{\mu\nu} q$ & $1/\Mstar^2$ & 6 \\
\Op{D11} & $\xbar  \x G_{\mu\nu}G^{\mu\nu}$ & $\alpha_S/4\Mstar^3$ & 7 \\
\Op{D12} & $\xbar \g5 \x G_{\mu\nu}G^{\mu\nu}$ & $i \alpha_S/4\Mstar^3$ & 7 \\
\Op{D13} & $\xbar  \x G_{\mu\nu}\tilde{G}^{\mu\nu}$ & $\alpha_S/4\Mstar^3$ & 7 \\
\Op{D14} & $\xbar \g5 \x G_{\mu\nu}\tilde{G}^{\mu\nu}$ & $i\alpha_S/4\Mstar^3$ & 7 \\
\hline
\end{tabular*}
  \caption{Operators for Dirac DM.}
  \label{DList}  
\end{table*}

\begin{table*}
  \begin{tabular*}{\textwidth}{@{\extracolsep{\fill}}lrrc}
  \hline
Label & \multicolumn{1}{c}{Operator} & \multicolumn{1}{c}{Usual coefficient} & Dimension \\
 \\
\hline
\Op{M1} & $\xbar  \x \qbar q$ & $m_q/2\Mstar^3$ & 6 \\
\Op{M2} & $\xbar i \g5 \x \qbar q$ & $m_q/2\Mstar^3$ & 6 \\
\Op{M3} & $\xbar   \x   \qbar i \g5  q$ & $m_q/2\Mstar^3$ & 6 \\
\Op{M4} & $\xbar i \g5 \x \qbar i \g5 q$ & $m_q/2\Mstar^3$ & 6 \\
\Op{M5} & $\xbar \gmu \g5 \x \qbar \gmd q$ & $1/2\Mstar^2$ & 6 \\
\Op{M6} & $\xbar \gmu \g5 \x \qbar \gmd \g5 q$ & $1/2\Mstar^2$ & 6 \\
\Op{M7} & $\xbar  \x G_{\mu\nu}G^{\mu\nu}$ & $\alpha_S/8\Mstar^3$ & 7 \\
\Op{M8} & $\xbar \g5 \x G_{\mu\nu}G^{\mu\nu}$ & $i \alpha_S/8\Mstar^3$ & 7 \\
\Op{M9} & $\xbar  \x G_{\mu\nu}\tilde{G}^{\mu\nu}$ & $\alpha_S/8\Mstar^3$ & 7 \\
\Op{M10} & $\xbar \g5 \x G_{\mu\nu}\tilde{G}^{\mu\nu}$ & $i\alpha_S/8\Mstar^3$ & 7 \\
\hline
\end{tabular*}
  \caption{Operators for Majorana DM. }
  \label{MList}  
\end{table*}

\begin{table*}
  \begin{tabular*}{\textwidth}{@{\extracolsep{\fill}}lrrc}
  \hline
Label & \multicolumn{1}{c}{Operator} & \multicolumn{1}{c}{Usual coefficient} & Dimension \\
 \\
\hline
\Op{C1} & $\phi^*  \phi  \qbar q$ & $m_q/\Mstar^2$ & 5 \\
\Op{C2} & $\phi^*  \phi  \qbar i \g5 q$ & $m_q/\Mstar^2$ & 5 \\
\Op{C3} & $\phi^* i\overleftrightarrow{\partial_\mu} \phi  \qbar  \gmu  q$ & $1/\Mstar^2$ & 6 \\
\Op{C4} & $\phi^* i\overleftrightarrow{\partial_\mu} \phi  \qbar \gmu \g5 q$ & $1/\Mstar^2$ & 6 \\
\Op{C5} & $\phi^*  \phi  G_{\mu\nu}G^{\mu\nu}$ & $\alpha_S/4\Mstar^2$ & 6 \\
\Op{C6} & $\phi^*  \phi  G_{\mu\nu}\tilde{G}^{\mu\nu}$ & $\alpha_S/4\Mstar^2$ & 6 \\
\hline
\end{tabular*}
  \caption{Operators for Complex Scalar DM. }
  \label{CList}  
\end{table*}

\begin{table*}
  \begin{tabular*}{\textwidth}{@{\extracolsep{\fill}}lrrc}
  \hline
Label & \multicolumn{1}{c}{Operator} & \multicolumn{1}{c}{Usual coefficient} & Dimension \\
 \\
\hline
\Op{R1} & $\phi^2  \qbar q$ & $m_q/2\Mstar^2$ & 5 \\
\Op{R2} & $\phi^2  \qbar i \g5 q$ & $m_q/2\Mstar^2$ & 5 \\
\Op{R3} & $\phi^2  G_{\mu\nu}G^{\mu\nu}$ & $\alpha_S/8\Mstar^2$ & 6 \\
\Op{R4} & $\phi^2  G_{\mu\nu}\tilde{G}^{\mu\nu}$ & $\alpha_S/8\Mstar^2$ & 6 \\
\hline
\end{tabular*}
  \caption{Operators for Real Scalar DM.}
    \label{RList}
\end{table*}

Generically, EFTs are a valid description of DM interactions with the Standard Model if the interactions are mediated by a heavy particle out of the kinematic reach of the collider. At the energy scales and coupling strengths accessible to the LHC, the validity of the EFT approximation can no longer be guaranteed.

As an illustration of the range of validity of EFT operators, we begin with a benchmark simplified model, where a pair of Dirac DM fermions interact with the SM via $s$-channel exchange of a $Z'$-like mediator with pure vector couplings 
\begin{equation}
\mathcal{L}_{\rm int} \supset -Z'_\mu( \sum_q \, \gq   \bar q \gamma^\mu  q + \gDM  \bar \chi \gamma^\mu\chi).
\label{lagrZprime}
\end{equation}
which is going to be discussed in detail in Sect.~\ref{subsec:1s12}.
The mediator has
mass $\mMed$ and vector couplings to quarks and DM with strength \gq and \gDM respectively, and this model reduces to the \Op{D5} operator in the full EFT limit.  At low energies, much smaller than $\mMed$, the heavy mediator can be integrated out 
and one is left with a theory without the mediator, where the interactions between DM and quarks
are described by a tower of effective operators.
The expansion in terms of this tower can be viewed
as the expansion of the  propagator of the mediator particle,
\be
\frac{g_q g_\chi}{\mMed^2-Q_{\rm tr}^2}
=\frac{g_q g_\chi}{\mMed^2}\left(1+\frac{Q_{\rm tr}^2}{\mMed^2}+\mathcal{O}\left(\frac{Q_{\rm tr}^4}{\mMed^4}\right)\right)\,,
\label{expansionEFT}
\ee
where \Qtr is the transfer momentum of the process.
Retaining  only the leading  term $1/\mMed^2$ corresponds to truncating the expansion to the 
lowest-dimensional operator. The parameters of the high-energy theory and the scale $\Mstar$ associated with the dimension-6 operators of the low-energy EFT are then connected via
\be
\Mstar=\frac{\mMed}{\sqrt{g_q g_\chi}},
\label{matching}
\ee
which holds as long as
\be
\Qtr \ll \mMed.
\label{ValidityCondition}
\ee

In such an $s$-channel  model, there is a condition defining the point where  the approximation has inevitably broken down.
The mediator must carry at least enough energy to produce DM at rest, therefore $Q_{\rm tr}>2m_{\rm DM}$.  Combining this with Eqs.~(\ref{matching})-(\ref{ValidityCondition}), we see
\be
\Mstar > \frac{Q_{\rm tr}}{\sqrt{g_q g_\chi}} > 2 \frac{m_{\rm DM}}{\sqrt{g_q g_\chi}}\,,
\label{kinconstraint}
\ee
which in the extreme case in which  couplings are as large as possible while remaining in the perturbative regime, $g_\chi,g_q<4\pi$, gives
\be
\Mstar>\frac{m_{\rm DM}}{2\pi}\,.
\label{mover2pi}
\ee
Note that this condition is necessary but not sufficient for the validity of the EFT approximation. A better measure of the validity comes from drawing a comparison between $\Qtr$ and $\mMed$, which defines three  regions \cite{1308.6799}:
\begin{enumerate}

\item When $\Qtr^2 < \mMed^2 \equiv \gq \gDM {\Mstar^2}$, the approximation in Eq.~(\ref{expansionEFT}) holds. This is clearly the only region where the EFT approximation  remains valid.

\item In the region where $\Qtr^2 \sim \mMed^2$ the production cross-section undergoes a resonant enhancement. The EFT approximation misses this enhancement, and is therefore conservative relative to the full theory.

\item When $\Qtr^2 \gg \mMed^2$, the expansion in Eq.~(\ref{expansionEFT}) fails and the signal cross section falls like $\Qtr^{-1}$ rather than $\mMed^{-1}$. In this region the EFT constraints will be stronger than the actual ones.

\end{enumerate}
Ref.~\cite{1507.00966} has calculated the kinematic distribution of events at 14 TeV for both this benchmark simplified model at a range of mediator masses, and the \Op{D5} operator. They find that the spectra become equivalent at a mediator mass of 10 TeV, and so EFTs can be considered a valid description of simplified models with mediators at or above this mass scale. At such large mediator mass scales, it is possible that a constraint on \Mstar will correspond to very large values of \gDM \gq above the range where perturbative calculations are valid. In this case it remains problematic to draw a clear correspondence between a constraint on \Mstar and a constraint on simplified model parameters.

EFTs do not aim to capture the complex physics described by UV-complete models, and so gauge invariance is often not enforced. 
This can lead to issues if the phenomenology of the operator no longer describes that of a UV complete operator but rather is symptomatic of the violation of gauge invariance. 
As an example, both ATLAS and CMS have included searches \cite{1309.4017,1407.7494,1408.2745} for a version of \Op{D5} where the relative coupling strength to up and down quarks was allowed to vary, leading to an enhancement of the cross section. Ref.~\cite{1503.07874} pointed out that this enhancement is due to the breaking of gauge invariance. In UV complete models that satisfy gauge invariance, the enhancement is much smaller \cite{1512.00476,1603.01267}.

Another issue that may arise when dealing with high-energy collisions is to make sure that unitarity of the S-matrix is not violated. When adopting an EFT description, this means that the condition of unitarity preservation sets an energy scale above which the contact interaction is not reliable anymore and a UV completion of the operator must be adopted instead.
For instance, for the operator \Op{D5}, the unitarity constraint gives
\cite{1112.5457}
\be
\Mstar > \left[\left(1-\frac{4\mDM^2}{s} \right) s  \frac{\sqrt{3}}{4 \pi} \right]^{1/2},
\ee
where $\sqrt{s}$ is center-of-mass energy of the initial state of the process $q\bar q\to \chi\bar\chi +j$
(see Ref.~\cite{1403.6610} for the constraints on other operators).
As a consistency check, the limits on \Mstar derived experimentally according with any of the two methods described below need to be compared with the unitarity bound.

\paragraph{\bf{EFT truncation by comparison with a simplified model.}} 
We see from Eqs.~(\ref{expansionEFT})-(\ref{matching}) that the validity of the EFT approximation as a description of some UV-complete model depends on the unknown parameters of that model. By introducing a minimum set of free parameters from such a model, one can enforce EFT validity by restricting the signal so that only events which pass the EFT validity condition Eq.~(\ref{ValidityCondition}) are used, thereby removing events for which the high-mediator-mass approximation made in the EFT limit is not a valid approximation in a given model.
In a typical $s$-channel model this EFT validity condition is 
\be
\Qtr^2 < \mMed^2 = \gq \gDM \Mstar^2.
\label{schanValidityCondition}
\ee
Discarding events which do not pass this condition gives a truncated signal cross section as a function of $(\mDM, \gq\gDM,  \Mstar)$ or $(\mDM, \mMed)$. This can be solved to find a rescaled, conservative limit on the energy scale, $\Mstar^{\rm rescaled}$. 

Note that if $\gq\gDM$ is fixed rather than $\mMed$, then the truncated cross-section which is used to derive a rescaled limit $\Mstar^{\rm rescaled}$ is itself a function of the $\Mstar^{\rm rescaled}$. Therefore the $\Mstar^{\rm rescaled}$ is found via a scan or iterative procedure. ATLAS has applied this procedure for a range of operators in Ref.~\cite{1502.01518}.

If instead $\mMed$ is fixed, then $\gq\gDM$ must increase to match the new value of $\Mstar$ via the relation in Eq.~(\ref{matching}). If a very large value of $\mMed$ is chosen or assumed in order to guarantee $\Qtr^2 < \mMed^2$, then the derived constraint on \Mstar may give a large value of $\gq \gDM$. If $\gq \gDM$ becomes sufficiently large, then perturbation theory is no longer a reliable computation technique.

\paragraph{\bf{EFT truncation using the center of mass energy.}} The procedure described above implicitly assumes some kind of knowledge of the underlying UV completion of the EFT. The truncation method relies on the transferred momentum \Qtr of the process of interest. 

Alternatively, it is possible to extract limits  without explicit assumptions about the UV completion, basing the truncation upon the center of mass energy $E_{\rm cm}$ of the process of DM production \cite{1502.04701,1402.7074}. 
The results will be more model-independent, but necessarily weaker than those based on the previous truncation method.

According to this method, the EFT approximation is reliable as long as
\begin{equation}
\label{Ecm<Mcut}
E_\text{cm}<M_\text{cut}\,,
\end{equation}
where the cutoff scale \Mcut is what defines the range of validity of the EFT approximation. Such scale can be related to the suppression scale \Mstar of the effective operator by $\Mcut=\gstar \, \Mstar\,$, where \gstar plays the role of an effective coupling, inherited by an unknown UV completion. For instance, in the case of a UV completion of the type \Zprime-type model of Eq.~\ref{lagrZprime}, one has  $\gstar=\sqrt{\gDM\gq}$.

As said, the parameter \Mcut is associated to the failure of the EFT description and it can be identified by using a ratio $R$, defined as the fraction of events satisfying the condition $\hat s<\Mcut^2$. Large enough \Mcut means all events are retained, so $R=1$. Small enough \Mcut means all events are rejected, so $R=0$, which means no result can be extracted. A useful methodology is to find the values of \Mcut for which the truncation provides values of $R$ within 0.1 and 1, and then show the corresponding limits for such values of \Mcut.

If a specific UV completion of the EFT is assumed (or hinted by experiments), the parameters \Mcut, \Mstar can be computed in terms of the paramters of the simplified model and the resulting bounds will be more conservative than those obtained by using \Qtr. However, if no UV completion is known or assumed, the method described here becomes particularly helpful.

In Ref.~\cite{ATL-PHYS-PUB-2014-007} the reader can find 
the details of an explicit application of these two truncation techniques.

\subsection{Effective Field Theories for direct detection\label{sec:eft-dd}}

Direct detection experiments search for the signature of DM scattering with a terrestrial target. Currently the most sensitive experiments use a  noble liquid target material in a two-phase time projection chamber. This design allows the experiment to see two signals: the prompt photons from scintillation events, and a delayed signal from ionisation events. The ratio between these two signals allows the experiment to distinguish between nuclear and electronic recoils, reducing the background from scattering due to cosmic rays and background radiation. This gives a constraint on the energy spectrum of DM-nucleus recoil events $dR/dE_R$, which is in turn used to constrain the DM-nucleon scattering cross-section via the relation (per unit target mass)
\be
\frac{dR}{dE_R} = \frac{\rho_\chi}{\mDM m_N}\int_{|\vec{v}| > v_{\rm min}} d^3 \vec{v} |\vec{v}| f(\vec{v})\frac{d\sigma_{\chi A}}{dE_R},
\ee
where $\rho_\chi$ is the local DM density, $v_{\rm min}=\sqrt{m_A E_R^{\rm th}/(2\mu_{\chi A})^2}$ is the minimum DM velocity required to transfer a threshold recoil kinetic energy $E_R^{\rm th}$ to the nucleus $A$, $\mu_{\chi A}$ is the DM-nucleus reduced mass, $f(\vec{v})$ is the local DM velocity distribution, and $d\sigma_{\chi A}/dE_R$ is the differential DM-nucleus scattering cross section. The energy dependence of  $d\sigma_{\chi A}/dE_R$ for a given detector depends on the underlying DM model and contains a nuclear form factor. This cross section can be computed starting from a more basic quantity, the DM-nucleon scattering cross section at zero momentum transfer $\sigma_{\chi N}$ (with $N=n,p$) which is the quantity commonly constrained by the experimental collaborations and can be thought of as the normalisation of the full cross-section $d\sigma_{\chi A}/dE_R$.

The scattering interactions involved in  direct detecion experiments are at a vastly different energy scale than those at the LHC. In a DM-nucleon scattering event, the DM velocity is of order $10^{-3} c$ and the momentum transfer is only $\mathcal{O}(10 \MeV)$ \cite{1008.1591}, which leads to two main differences when compared with the picture at colliders: (1) in direct detection experiments, the EFT approximations will be valid for a much larger range of parameters, down to mediators at the MeV mass scale \cite{1412.6220}; and (2) the relevant operators are not the usual DM-parton operators considered in Section~\ref{sec:eft-lhc}, but rather the non-relativistic limit of DM-nucleon operators. A partial list of these operators is given in Table~\ref{NRList}, in the language of \cite{1203.3542,Cirelli}. 
The discussion in this section is limited to the matching of operators at the lowest order. For long-distance next-to-leading-order QCD corrections to WIMP-nucleus cross section see e.g. Refs.~\cite{astro-ph/0309115,1205.2695,1208.1094,1304.7684,1309.0825,1311.5886,1408.5046,1412.6091,1503.04811}.

The large splitting between the LHC and direct detection energy scales makes it important to remember that the operator coefficients need to be RG-evoluted from the high energy theory, including the matching conditions at the quark masses thresholds \cite{1402.1173,1411.3342}. 

The matrix element describing DM-nucleon contact interactions is then given by a sum of the contributions from each non-relativistic operator
\be
\mathcal{M} = \sum_{i=1}^{12} \cgoth^N_i(\mDM) \Op[NR]{i}.
\label{NRMatrixElement}
\ee
\begin{table*}
\centering
  \begin{tabular*}{0.4\textwidth}{@{\extracolsep{\fill}}lr@{}}
  \hline
Label & \multicolumn{1}{c}{Operator}  \\
 \\
\hline
\Op[NR]{1} & $\mathbbm{1}$ \\
\Op[NR]{3} & $i \vec{s}_N \cdot (\vec{q} \times \vec{v}^\bot)$  \\
\Op[NR]{4} & $ \vec{s}_\chi \cdot \vec{s}_N $  \\
\Op[NR]{5} & $ i \vec{s}_\chi \cdot (\vec{q} \times \vec{v}^\bot) $  \\
\Op[NR]{6} & $ (\vec{s}_\chi \cdot \vec{q})(\vec{s}_N \cdot \vec{q})$  \\
\Op[NR]{7} & $\vec{s}_N \cdot \vec{v}^\bot$  \\
\Op[NR]{8} & $\vec{s}_\chi \cdot \vec{v}^\bot$  \\
\Op[NR]{9} & $i\vec{s}_\chi \cdot (\vec{s}_N \times \vec{q})$  \\
\Op[NR]{10} & $i \vec{s}_N \cdot \vec{q} $  \\
\Op[NR]{11} & $ i \vec{s}_\chi \cdot \vec{q} $  \\
\Op[NR]{12} & $ \vec{v}^\bot \cdot (\vec{s}_\chi \times \vec{s}_N)$  \\
\hline
\end{tabular*}
  \caption{Non-relativistic DM-nucleon contact operators relevant to describing the interactions listed in Section~\ref{sec:eft-lhc}.
  The operator $\Op[NR]{2}=( \vec{v}^\bot)^2$ from Ref.~\cite{1203.3542} is not induced by any of the relativistic operators considered in Sec.~\ref{sec:eft-lhc} and so is not discussed here.}
  \label{NRList}  
\end{table*}
Next we show how to translate between the language of relativistic DM-quark operators discussed in Section~\ref{sec:eft-lhc} and direct detection constraints on the non-relativistic DM-nucleon operators in Table~\ref{NRList} \cite{1008.1591,1203.3542,Cirelli}. 

To do this, first we consider the intermediate-stage relativistic DM-nucleon operators, beginning with the Dirac DM listed in Table~\ref{DMNucleonList} as a concrete example, with other cases discussed later. 
The effective Lagrangian at nucleon level gains contributions from  DM interactions with quarks and gluons and can be written at either level as
\be
\mathcal{L}_{\rm eff} = \sum_{q,i} c^q_i \Op[q]{i}+\sum_{g,j} c^g_i \Op[g]{j} =  \sum_{N,k} c^N_k \Op[N]{k},
\label{lagrangian-equivalence}
\ee
where $i,j$ are summed over whichever operators are present in the model of interest, and $N=n,p$. This will induce a sum over some subset $k$ of nucleonic operators. 
The value of the coefficients $c^N_k$, given in the third column of Table~\ref{DMNucleonList}, are a function of the coefficients of the DM-quark and DM-gluon operators, $c^q_i$ and $c^g_j$. These are dimensionful coefficients, with the usual parameterisation given in the third column of Table~\ref{DList} for Dirac DM.

The coefficients $c^N_k$ in Table~\ref{DMNucleonList} are also a function of several other parameters. Note that $f^{(N)}_{G}\equiv 1- \sum_{q=u,d,s} f^{(N)}_{q}$, and $C_{3,4}=\left( \sum_q {c_{3,4}^q}/{m_q}\right)\left(\sum_{q=u,d,s}m_q^{-1}\right)^{-1}$.
There is some uncertainty in the determination of $f^{(N)}_{q}$, $\delta^{(N)}_q$ and $\Delta^{(N)}_q$. 
In Table~\ref{DDConstants}, we show the values used by {\tt micrOMEGAs} \cite{1305.0237}. Although they use a relatively old determination of these parameters, they remain useful as a benchmark commonly used by the community. 
 Note that other, quite different sets of values are also available  in the literature. 
See  Refs.~\cite{hep-ph/0001005, astro-ph/0406204, 0801.3656, 0803.2360, 1110.3797,1202.1292,1209.2870,1312.4951,1506.04142,1510.06039}
for $f_{q}^{(N)}, \Delta_q^{(N)}$
and Refs.~\cite{0812.4366, 1303.3822,1212.3568} for other determinations of $\delta^{(N)}_{q}$.


\begin{table*}
  \begin{tabular*}{\textwidth}{@{\extracolsep{\fill}}lllll@{}}
  \hline
Label & \multicolumn{1}{c}{Operator} & \multicolumn{1}{c}{DM-parton coefficient $c_k^N$} \\
 \\
\hline
\Op[N]{D1} & $\xbar  \x \bar{N} N$ & $\sum_{q=u,d,s} c^q_{\rm D1} \frac{m_N}{m_q} f^{(N)}_{q} + \frac{2}{27}f^{(N)}_{G}( \sum_{q=c,b,t} c^q_{\rm D1} \frac{m_N}{m_q} - \frac{1}{3 \pi}c_{\rm D11}^g m_N)$ \\
\Op[N]{D2} & $\xbar i \g5 \x \bar{N} N$ & $\sum_{q=u,d,s} c^q_{\rm D2} \frac{m_N}{m_q} f^{(N)}_{q} + \frac{2}{27}f^{(N)}_{G} ( \sum_{q=c,b,t} c^q_{\rm D2} \frac{m_N}{m_q} - \frac{1}{3  \pi}c_{\rm D12}^g m_N)$ \\
\Op[N]{D3} & $\xbar   \x   \bar{N}  i \g5  N$ & $\sum_{q=u,d,s} \frac{m_N}{m_q} [(c^q_{\rm D3} - C_3)+\frac{1}{2  \pi}c^g_{\rm D13} \tilde{m} ] \Delta^{(N)}_q$\\
\Op[N]{D4} & $\xbar i \g5 \x \bar{N} i \g5 N$ & $\sum_{q=u,d,s} \frac{m_N}{m_q} [(c^q_{\rm D4} - C_4)+\frac{1}{2  \pi}c^g_{\rm D14} \tilde{m} ] \Delta^{(N)}_q$\\
\Op[N]{D5} & $\xbar \gmu \x \bar{N} \gmd N$ & $2c^u_{\rm D5}+c^d_{\rm D5}$ for \Op[p]{D5}, and $c^u_{\rm D5}+2c^d_{\rm D5}$ for \Op[n]{D5} \\
\Op[N]{D6} & $\xbar \gmu \g5 \x \bar{N} \gmd N$ & $2c^u_{\rm D6}+c^d_{\rm D6}$ for \Op[p]{D6}, and $c^u_{\rm D6}+2c^d_{\rm D6}$ for \Op[n]{D6}\\
\Op[N]{D7} & $\xbar \gmu \x \bar{N} \gmd \g5 N$ & $\sum_q c^q_{\rm D7} \Delta^{(N)}_q$\\
\Op[N]{D8} & $\xbar \gmu \g5 \x \bar{N} \gmd \g5 N$ & $\sum_q c^q_{\rm D8} \Delta^{(N)}_q$\\
\Op[N]{D9} & $\xbar \sigma^{\mu \nu} \x \bar{N} \sigma_{\mu\nu} N$ & $\sum_q c^q_{\rm D9} \delta^{(N)}_q$\\
\Op[N]{D10} & $\xbar i \sigma^{\mu \nu} \g5 \x \bar{N} \sigma_{\mu\nu} N$ & $\sum_q c^q_{\rm D10} \delta^{(N)}_q$ \\
\hline
\end{tabular*}
  \caption{DM-nucleon operators for Dirac fermion DM. For Majorana DM, \Op{D5}, \Op{D7}, \Op{D9} and \Op{D10} disappear. The coefficients $c^q_{\rm D1 ... D10}$, $c^g_{\rm D11 ... D13}$ are the corresponding coefficients from the third column of Table~\ref{DList}, e.g. $c^q_{\rm D5} = 1/\Mstar^2$. Recall that the coefficients are in principle independent for each quark flavor. }
   \label{DMNucleonList} 
\end{table*}

\begin{table*}
  \begin{tabular*}{\textwidth}{@{\extracolsep{\fill}}ccccccccc @{}}
  \hline
$f^{(p)}_{u}$ & $f^{(p)}_{d}$ & $f^{(p)}_{s}$ & $\Delta^{(p)}_u$ & $\Delta^{(p)}_d$  & $\Delta^{(p)}_s$ & $\delta^{(p)}_u $ & $\delta^{(p)}_d$ & $\delta^{(p)}_s$   \\
\hline
0.0153 & 0.0191 & 0.0447 & 0.842 & -0.427 & -0.085 & 0.84 & -0.23 & -0.046  \\
\hline
\end{tabular*}
  \begin{tabular*}{\textwidth}{@{\extracolsep{\fill}}ccccccccc @{}}
\hline
$f^{(n)}_{u}$ & $f^{(n)}_{d}$ & $f^{(n)}_{s}$ & $\Delta^{(n)}_u$ & $\Delta^{(n)}_d$  & $\Delta^{(n)}_s$ & $\delta^{(n)}_u $ & $\delta^{(n)}_d$ & $\delta^{(n)}_s$   \\
\hline
0.011 & 0.0273 & 0.0447 &  -0.427 & 0.842 & -0.085 & -0.23 & 0.84 & -0.046  \\
\hline
\end{tabular*}
  \caption{Quark-nucleon form factors as used by {\tt micrOMEGAs} \cite{1305.0237}. Note that  $f^{(p)}_{s}= f^{(n)}_{s}$, $\Delta^{(p)}_u = \Delta^{(n)}_d$, $\Delta^{(p)}_d = \Delta^{(n)}_u$, etc.   }
   \label{DDConstants} 
\end{table*}

The next step is to establish relationships between relativistic and non-relativistic operators.
At leading order in the non-relativistic limit, the DM-nucleon operators in Table~\ref{DMNucleonList} reduce down to a combination of the operators from Table~\ref{NRList} according to the relations
\begin{eqnarray}
\langle \Op[N]{D1} \rangle =  \langle \Op[N]{D5} \rangle & = & 4\mDM m_N \Op[NR]{1}, \nonumber \\
\langle \Op[N]{D2} \rangle & = & -4m_N\Op[NR]{11}, \nonumber \\
\langle \Op[N]{D3} \rangle & = &4\mDM \Op[NR]{10}, \nonumber \\
\langle \Op[N]{D4} \rangle & = &4\Op[NR]{6} , \nonumber \\
\langle \Op[N]{D6} \rangle & = &8\mDM(m_N\Op[NR]{8} + \Op[NR]{9}), \nonumber \\
\langle \Op[N]{D7} \rangle & = &8m_N(-\mDM\Op[NR]{8} + \Op[NR]{9}), \nonumber \\
\langle \Op[N]{D8} \rangle = -\frac{1}{2}\langle \Op[N]{D9} \rangle & = &-16 \mDM m_N \Op[NR]{4}, \nonumber \\
\langle \Op[N]{D10} \rangle & =& 8(\mDM \Op[NR]{11} -m_N \Op[NR]{10} - 4\mDM m_N \Op[NR]{12}.
\label{NRNRelationsD}
\end{eqnarray}
Using these relationships, the matrix-element for the interactions described by the Lagrangian in Eq.~(\ref{lagrangian-equivalence}) can be rewritten in terms of a sum of non-relativistic operators. Used in combination with Eq.~(\ref{NRMatrixElement}), the coefficients $\cgoth_i^N$ of the NR operators can be converted into those of the relativistic operators and vice-versa.

As an example, let us consider the \Op{D5} operator. If  the coupling to each flavor of quark is chosen to be independent, i.e. $c^q_{\rm D1} = 1/M_{*,q}^2$, then the effective Lagrangian at the DM-quark level is 
\be
\mathcal{L}_{\rm eff} = \sum_{q} c^q_{\rm D5} \Op[q]{D5} =\sum_{q} \frac{1}{M_{*,q}^2} \xbar \gmu \x \bar q \gmd q.
\ee
Combining this with the information in Table~\ref{DMNucleonList}, we see that this operator contributes to $\Op[N]{D5}$, and so the effective Lagrangian at DM-nucleon level is
\begin{eqnarray}
\mathcal{L}_{\rm eff} &= \sum_{N} c^N_{\rm D5} \Op[N]{D5} &= \left( 2 c^u_{\rm D5} + c^n_{\rm D5} \right) \Op[p]{D5} + \left(  c^u_{\rm D5} + 2 c^n_{\rm D5} \right) \Op[n]{D5} \nonumber\\
&&= \left( \frac{2}{M_{*,u}^2}  + \frac{1}{M_{*,d}^2} \right)  \xbar \gmu \x \bar{p} \gmd p + \left( \frac{1}{M_{*,u}^2}  + \frac{2}{M_{*,d}^2} \right)  \xbar \gmu \x \bar{n} \gmd n.
\end{eqnarray}
Using Eq.~(\ref{NRNRelationsD}), we see that $\langle \Op[N]{D5}\rangle = 4\mDM m_N \Op[NR]{1}$, therefore the matrix element is
\begin{eqnarray}
\mathcal{M} &=& \sum_{N} c^N_{D5} \langle\Op[N]{D5}\rangle\nonumber\\
& =& \left[4\mDM m_p \left(\frac{2}{M_{*,u}^2}  + \frac{1}{M_{*,d}^2} \right) + 4\mDM m_n \left(\frac{1}{M_{*,u}^2}  + \frac{2}{M_{*,d}^2} \right) \right] \Op[NR]{1}\nonumber\\
&=&\left(\cgoth^p_1(\mDM)+\cgoth^n_1(\mDM)\right) \Op[NR]{1}
\end{eqnarray}
Ref.~\cite{Cirelli} provides a toolset to convert experimental data into a constraint on any combination of relativistic or non-relativistic operators, by defining a benchmark constraint on an arbitrary operator and using the  conversion formula
\be
\cgoth^p_i(\mDM)^2 = \sum_{i,j=1}^{12} \sum_{N,N'=p,n} \cgoth_i^{N}(\mDM) \cgoth_J^{N'}(\mDM) \mathcal{Y}_{i,j}^{(N,N')}(\mDM)
\label{CoreDDConversion}
\ee
where $\mathcal{Y}_{i,j}^{(N,N')}$ are given as a set of interpolating functions for each experiment. 

To reiterate, in this section we have summarised how to convert between the coefficients $\cgoth_i^N(\mDM)$ of the NR operators relevant for direct detection, and the coefficients $c_i^q$ or $\Mstar$ of the fundamental underlying DM-parton operators. With this information, Eq.~(\ref{CoreDDConversion}) can be used to convert between constraints on different operators using e.g. the code given in Ref.~\cite{Cirelli}.

Moving beyond Dirac DM, the relationships between operators for Majorana DM are very similar to those given in Table~\ref{DMNucleonList}, the difference being that \Op{D5}, \Op{D7}, \Op{D9} and \Op{D10} disappear and so do not have a Majorana analogue. Therefore \Op[N]{D5}, \Op[N]{D7} also do not have a Majorana version. 

For complex scalar DM, the DM-nucleon operators are given in Table~\ref{ScalarDMNucleonList}.
\begin{table*}
  \begin{tabular*}{\textwidth}{@{\extracolsep{\fill}}lllll@{}}
  \hline
Label & \multicolumn{1}{c}{Operator} & \multicolumn{1}{c}{DM-parton coefficient  $c_k^N$} \\
 \\
\hline
\Op[N]{C1} & $\phi^* \phi \bar{N} N$ & $\sum_{q=u,d,s} c^q_{\rm C1} \frac{m_N}{m_q} f^{(N)}_{q} + \frac{2}{27}f^{(N)}_{G}( \sum_{q=c,b,t} c^q_{\rm C1} \frac{m_N}{m_q} - \frac{1}{3 \pi }c^g_{\rm C5} m_N)$ \\
\Op[N]{C2} & $\phi^* \phi \bar{N} i \gamma_5 N$ & $\sum_{q=u,d,s} \frac{m_N}{m_q} [(c^q_{\rm C2} - C_2)+\frac{1}{2 \pi }c^g_{\rm C6} \tilde{m} ] \Delta^{(N)}_q$\\
\Op[N]{C3} & $\phi^* i\overleftrightarrow{\partial_\mu} \phi  \bar{N}  \gmu  N$ & $2c^u_{\rm C3}+c^d_{\rm C3}$ for \Op[p]{C3}, and $c^u_{\rm C3}+2c^d_{\rm C3}$ for \Op[n]{C3} \\
\Op[N]{C4} & $\phi^* i\overleftrightarrow{\partial_\mu} \phi  \bar{N}  \gmu \g5  N$ & $\sum_q c^q_{\rm C4} \Delta^{(N)}_q$ \\
\hline
\end{tabular*}
  \caption{DM-nucleon operators for complex scalar fermion DM.}
  \label{ScalarDMNucleonList}  
\end{table*}
At leading order in the non-relativistic limit,   these reduce to
\begin{eqnarray}
\langle \Op[N]{C1} \rangle & = & 2m_N\Op[NR]{1}, \nonumber \\
\langle \Op[N]{C2} \rangle & = & 2\Op[NR]{10}, \nonumber \\
\langle \Op[N]{C3} \rangle & = & 4\mDM m_N\Op[NR]{1} , \nonumber \\
\langle \Op[N]{C4} \rangle & = & -8\mDM m_N\Op[NR]{7}. 
\label{NRNRelationsC}
\end{eqnarray}
For real scalar DM, $\phi^* \equiv \phi$ and \Op[N]{C3}, \Op[N]{C4} vanish.

The final step to make contact with experimental results is to draw a relationship between the coefficients of the DM-parton operators and the notation used in the direct detection community, where constraints on the scattering rate are usually given in terms of either spin-independent scattering cross section \sigSI, or the spin-dependent scattering cross section \sigSD.
These two parameterisations of the scattering rate are induced by the lowest-order expansion of specific non-relativistic operators. 

The spin-independent scattering rate corresponds to a constraint on $\cgoth^N_1$ of \Op[NR]{1}. This operator is the only one not suppressed by either the momentum of the DM or a spin coupling, and so is the most commonly studied interaction in the community. 

The spin-dependent rate \sigSD corresponds to a constraint on $\cgoth^N_4$ of  \Op[NR]{4}. This corresponds to an interaction of the DM spin with the nuclear spin and therefore the scattering rate is suppressed by the spin of the target nucleus. Not all experiments are sensitive to this interaction. 

From Eqs.~(\ref{NRNRelationsD}) and (\ref{NRNRelationsC}) we see that several DM-nucleon operators lead to these two NR operators. Specifically: \Op[N]{D1},  \Op[N]{D5},  \Op[N]{C1}, and  \Op[N]{C3} lead to \Op[NR]{1}, while  \Op[N]{D8},  \Op[N]{D9} lead to  \Op[NR]{4}. At the DM-quark level, \Op{D1}, \Op{D5},  \Op{D11}, \Op{C1}, \Op{C3} and \Op{C5} each  lead to a spin-independent scattering cross section, while \Op{D8}, \Op{D9} lead to a spin-dependent scattering cross section. The formula for $\sigSI, \, \sigSD$ for each of these operators \Op{i} is
\begin{eqnarray}
\sigSI &=& \frac{\mu_{\rm \chi N}^2}{\pi} \left( c_i^N \right)^2,\label{sigSI}\\
&&\nn\\
\sigSD &=& \frac{3\mu_{\rm \chi N}^2}{\pi} \left( c_i^N \right)^2,\label{sigSD}
\end{eqnarray}
where $c_i^N$ is given in Table~\ref{DMNucleonList} for Dirac fermion DM and Table~\ref{ScalarDMNucleonList} for complex scalar DM,  $\mu_{\rm \chi N}=\mDM m_N/(\mDM + m_N)$ is the DM-nucleon reduced mass, and the target nucleon is either a neutron or a proton $N=n,p$.

The precise application of these formulae to convert between $\sigSI, \, \sigSD$ and the usual coefficients $c_i^q$ from Tables~\ref{DList}, \ref{CList} is sensitive to the choice of the nuclear form factors (see Table~\ref{DDConstants}), and so we list here the usual conversion used by the community \cite{1008.1783},
\begin{eqnarray}
\sigSI^{\rm D1} &=& 1.60 \times 10^{-37} \cm^2 \left( \frac{\mu_{\rm \chi,N}}{1 \GeV}\right)^2\left(\frac{20 \GeV}{\Mstar}\right)^6\\
\sigSI^{\rm D5,C3} &=& 1.38 \times 10^{-37} \cm^2 \left( \frac{\mu_{\rm \chi,N}}{1 \GeV}\right)^2\left(\frac{300 \GeV}{\Mstar}\right)^4\\
\sigSI^{\rm D11} &=& 3.83 \times 10^{-41} \cm^2 \left( \frac{\mu_{\rm \chi,N}}{1 \GeV}\right)^2\left(\frac{100 \GeV}{\Mstar}\right)^6\\
\sigSI^{\rm C1} &=& 2.56 \times 10^{-36} \cm^2 \left( \frac{\mu_{\rm \chi,N}}{1 \GeV}\right)^2\left(\frac{10 \GeV}{\Mstar}\right)^4\left(\frac{10 \GeV}{\mDM}\right)^2\\
\sigSI^{\rm C5} &= &7.40 \times 10^{-39} \cm^2 \left( \frac{\mu_{\rm \chi,N}}{1 \GeV}\right)^2\left(\frac{60 \GeV}{\Mstar}\right)^4\left(\frac{10 \GeV}{\mDM}\right)^2\\
\sigSD^{\rm D8,D9} &=& 4.70 \times 10^{-39} \cm^2 \left( \frac{\mu_{\rm \chi,N}}{1 \GeV}\right)^2\left(\frac{300 \GeV}{\Mstar}\right)^4.
\end{eqnarray}
It is possible to convert constraints on $\sigSI$ and $\sigSD$ into constraints on the parameters of any other operator or combination of operators using Eq.~(\ref{CoreDDConversion}) with the code described in Ref.~\cite{Cirelli}.

\subsection{Effective Field Theories for indirect detection\label{sec:eft-id}}

Indirect detection is the search for the Standard Model particles arising as a result of DM self-annihilations  (see e.g. Ref.~\cite{1511.02031} for a state-of-the-art review). DM annihilation takes place on many scales, from cosmological scales down to annihilation within the solar system. 

Most indirect detection studies search for the gamma-ray signal from WIMP annihilation on the scale of Galactic halos. Both direct production of photons and secondary production from the decay of other SM particles are considered. For annihilation of DM of mass $\mDM$ within the Galactic halo, the gamma-ray flux observed at Earth along a line of sight at angle $\psi$ from the Galactic center, with an initial photon energy spectrum per annihilation given by $dN_\gamma/dE$, reads

\be
\frac{d\Phi_{\gamma}}{dE} = \frac{1}{2}\frac{\sigv_{\rm total}}{4 \pi \mDM ^2} \frac{dN_\gamma}{dE} \frac{\mathcal{J}(\psi)}{{\rm J}_0},
\ee
where 
\be
\mathcal{J}(\psi) = {\rm J_0} \int^{\ell_{max}}_0 \rho^2\left(\sqrt{R_{\textrm{sc}}^2 -
2\ell R_{\textrm{sc}}\cos{\psi} +\ell^2}\right)d\ell\,,
\ee
is the integrated DM density squared and  J$_0=1/[8.5\,{\rm kpc} \times(0.3 \,{\rm GeV \, cm}^{-3})^2]$ is an arbitrary normalization constant used to make $\cal{J}(\psi)$ dimensionless.

This form of the expression is useful as it factorizes $\mathcal{J}$, which depends on astrophysics, from the rest of the expression which depends on particles physics. With knowledge of  $\mathcal{J}$ for the studied annihilation region and the gamma-ray spectrum per annihilation ${dN_\gamma}/{dE}$, a constraint can be placed on the thermally averaged self-annihilation cross-section, $\sigv_{\rm total}$. A constraint on this parameter depends only on the spectrum of SM particles per annihilation, not on the underlying particle physics model. The numerical tool introduced in Ref.~\cite{1012.4515} is helpful to get the spectrum of SM particles in the final state of DM annihilations.
Since this spectrum is unknown, searches typically present constraints on individual channels assuming 100\% branching ratio to that channel. For example, a search may present a constraint assuming annihilation purely to $W^+W^-$. This is equivalent to a constraint on the total cross section scaled by the branching ratio to that final state, $\sigv_{W^+W^-}\equiv\sigv_{\rm total}\times BR(W^+W^-)$. 

This means that an EFT analysis is not strictly necessary for indirect detection studies, since the calculation of the  branching ratios within a specific model only adds model-dependence to the constraints. 

There are specific cases where EFT can be useful, such as if one is interested in the spectrum of gamma-rays from DM annihilation taking into account all final states. For example, Ref.~\cite{1403.5027} used effective operators to study whether DM can produce the spectrum of a potential gamma-ray excess from the galactic center,
and Refs.~\cite{1009.0008,1211.7061,1303.4423,1409.8294,1205.4723} uses the EFT formalism to calculate the DM annihilation rate to the $\gamma\gamma$ final state. This is a very clean signature with few astrophysical backgrounds, and so determining an accurate branching ratio to this final state can give very strong constraints on DM models.

Effective operators are also useful as a way to compare the strength of indirect detection constraints with constraints from other searches such as direct detection experiments and colliders~\cite{1011.2310,1104.5329,1201.3402,1212.3990,1506.08841,1601.06140}.

Galactic WIMPs at the electroweak scale are non-relativistic, and so the energy scale of the interaction is of order $2\mDM$. Hence the EFT approximation is valid for indirect detection experiments as long as the DM mass is much lighter than the mediator mass.

The operators describing DM interactions with the SM can be organized 
in the non-relativistic limit as an expansion in their mass dimension and in their velocity dependence (e.g. $s$-wave, $p$-wave, etc. annihilations).
For self-conjugate DM (a Majorana fermion or a real scalar field),  DM annihilation to light fermions suffers from helicity suppression which can be lifted by including extra gauge boson radiation. This effect is of particular relevance for indirect detection, as it can significantly change the energy spectra of stable particles originating from DM annihilations \cite{Bergstrom:1989jr,0710.3169, 0808.3725, 1009.0224, 1009.2584, 1101.3357, 1104.2996, 1104.3823, 1104.5329, 1105.5367, 1107.4453, 1111.2916, 1111.4523, 1112.5155, 1201.1443}. 
Sticking to the EFT framework, this effect is encoded by higher-dimensional operators 
\cite{1301.1486,1601.06140}.

Indirect detection can also be used to constrain the WIMP-nucleon scattering rate via neutrinos from the sun. As the solar system passes through the Galactic DM halo, DM will scatter with the sun and become gravitationally bound. The DM annihilation rate depends on the square of the DM number density, and therefore after a sufficient amount of time has passed, the DM annihilation rate will increase until it reaches equilibrium with the scattering rate. Therefore the size of the scattering rate will control the flux of particles from the sun from DM annihilation. Due to the opacity of the sun, neutrinos are the only observable DM annihilation product from the sun, and so IceCube and other neutrino observatories can use limits on the neutrino flux from the sun to place constraints on the DM scattering cross-section \cite{1601.00653}.
This means that indirect detection is in the unique position of being able to probe both the relativistic and non-relativistic DM-SM effective operators.

%% file: sections/simp.tex
\section{A paradigm shift: Simplified Models}
\label{sec:simp}
\def\missET {\slashed{E}_T}

In the previous section we have spelled out the virtues and drawbacks of the
EFT approach for DM searches. Whilst the EFT remains a useful tool if used consistently, it is now clear that we must also look beyond the effective operator approximation.

As anticipated in the Introduction to this review, a possible alternative approach consists of expanding the contact interaction of DM with the SM and include the ``mediator'' as propagating degrees of freedom of the theory. By increasing the number of parameters necessary to specify the unknown DM interactions one gains a more complete theoretical control. 

In this section we will summarize the phenomenology of the simplified models for DM and, wherever available, provide the most important results concerning the collider searches, the DM self-annihilation cross sections and the cross sections for DM scattering with nucleons.

So far, as is customary when discussing EFTs, we have followed a bottom-up approach: the list of effective operators comes purely from symmetry and dimensional analyses. The shift to simplified models now makes it more advantageous  to reverse the logic and use a top-down approach from here on. We will categorize the models according to the quantum numbers of the DM particle and the mediator, and to the mediator type (s- or t-channel); see Table \ref{SimpModelsTable}. This classification refers to $2\rightarrow 2$ tree-level processes and the model names we choose are designed as an easy-to-recall nomenclature.

We have decided to limit the discussion to scalar and fermion DM only, and not to include in the list the cases where the DM is a massive vector particle.
In the spirit of the simplified models, the smallest possible number of degrees of freedom should be added to the SM. Also, the model building with vector DM is necessarily more involved.
Furthermore,  many DM searches at the LHC are based on counting analyses, for which the DM spin is typically not very relevant. Event topologies more complex than the $\missET+j$ can be constructed, along with angular variables \cite{1311.7131,1501.00907}, which would also allow the exploration of the spin of the DM particle, to some extent. However, we believe that at the present stage of LHC searches for DM, the simplified models discussed in this review already capture a very rich phenomenology.

Before reviewing the features and the phenomenology of all the cases listed in Table \ref{SimpModelsTable}, we first point out some general properties of simplified models.

\begin{table}
\begin{center}
\begin{large}
 \begin{tabular}{|c|c|c|c|c|}
  \hline
Mediator spin & Channel & DM spin &Model Name & Discussed in Section \\
\hline
0&$s$&0& $0s0$ & \ref{subsec:0s0} \\
&&&&\\
0&$s$&$\frac12$& $0s\frac12$ & \ref{subsec:0s12}\\
&&&&\\
0&$t$&$\frac12$& $0t\frac12$ & \ref{subsec:0t12} \\
&&&&\\
\hline
$\frac12$&$t$&0& $\frac12t0$ & \ref{subsec:12t0}\\
&&&&\\
$\frac12$&$t$& $\frac12$& $\frac12t\frac12$ & \ref{subsec:12t12}\\
&&&&\\
\hline
1&$s$&0& $1s0$ & \ref{subsec:1s0}\\
&&&&\\
1&$s$&$\frac12$& $1s\frac12$ & \ref{subsec:1s12}\\
&&&&\\
1&$t$&$\frac12$& $1t\frac12$ & \ref{subsec:1t12} \\
&&&&\\
\hline
\end{tabular}
  \caption{Simplified models for scalar and fermion DM.}
  \label{SimpModelsTable}
\end{large}
\end{center}  
\end{table}

\subsection{General properties of simplified models}

As discussed above, when building a simplified model for DM one wants to extend the SM by adding new degrees of freedom: not too many, otherwise simplicity is lost; not too few, otherwise the relevant physics is not described completely.
To this end, one builds simplified models according to the following general prescriptions:

\begin{itemize}
\item[(i)] the SM is extended by the addition of a DM particle, which is absolutely stable (or, at least, stable on collider scale).

\item[(ii)] The new Lagrangian operators of the models are renormalizable and consistent with the symmetries: Lorentz invariance, SM gauge invariance, DM stability.
\end{itemize}

In addition to these exact symmetries, the SM has other important global symmetries. Baryon and lepton number are anomalous, but they can be treated as exact symmetries at the renormalizable level. So, we require that simplified models respect baryon and lepton number.

On the other hand, the flavor symmetry of the SM can be broken by new physics, but we need to ensure this breaking is sufficiently small to agree with high-precision flavor experiments. One very convenient approach to deal with this is to impose that new physics either respect the SM flavor symmetry or the breaking of it is associated with the quark Yukawa matrices.

This idea is known as Minimal Flavor Violation (MFV) \cite{hep-ph/0207036}. 
It also allows us to keep small the amount of CP violating effects  which are possibly induced by new physics.
Throughout this paper we will adopt MFV, although it would be interesting to have results for simplified models also in the very constrained situations where this assumption is relaxed.

Following the guidelines outlined above, we now proceed to build and discuss the phenomenology of simplified models.

\subsection{Scalar Mediator}
\label{subsec:scalarmediator}

The simplest type of simplified model is the one where a scalar particle mediates the interaction between DM and the SM. Their interaction can occur via s-channel or t- channel diagrams. The scalar mediator could be real or complex. In the complex case, it has both scalar and pseudoscalar components. We will separately discuss  the cases where the mediator is a purely scalar or a purely pseudoscalar particle.

As for the DM, it may either be a scalar ($0s0$ model) or a Dirac or Majorana fermion ($0s\frac{1}{2}$ model). The more complex possibility of a fermion DM being a mixture of an EW singlet and doublet will be discussed later, as it leads to a hybrid $0s\frac{1}{2}/1s\frac{1}{2}$ model.

The primary focus will be on the tree-level mediator couplings to SM fermions (and the couplings  to gluons arising at one loop), being the most important for LHC phenomenology.

An important aspect to keep in mind when dealing with scalar mediators is that they generically mix with the neutral Higgs. In turn, this would affect the Yukawa couplings and the tree-level vertices of the Higgs with two gauge bosons.
Such deviations with respect to SM Higgs couplings are severely constrained by Higgs production and decay measurements, although not excluded completely.
A common approach in the literature is to simply set  the mixing of the scalar mediator with the Higgs to zero,  thus keeping the minimal  possible set of parameters.

On the other hand one must also consider the possibility that the Higgs boson itself can serve as a scalar mediator between the DM and the rest of the SM, thus providing a rather economical scenario in terms of new degrees of freedom, and sometimes a richer phenomenology. Connecting the DM sector to the SM via the Higgs field may have also interesting consequences for the electroweak symmetry breaking and the Higgs vacuum stability, and it is possible to link the solutions of the hierarchy problem and of the DM problem in a unified framework
\cite{1306.2329,1403.4953,1408.3429,ColemanWeinberg,hep-th/0612165,0709.2750,1301.4224,1506.03285}.

\subsubsection{Scalar DM, $s$-channel ($0s0$ model)}
\label{subsec:0s0}

In the case where DM is a real scalar singlet $\phi$, the mediation is via s-channel and the mediator is a neutral scalar. The most minimal choice is to consider the Higgs boson $h$ as a mediator, rather than a speculative dark sector particle \cite{hep-ph/0011335,0706.4311,1112.3299,1205.3169,1306.4710,1403.4953,1412.0258,1509.01765,1511.01099,1601.06232}.

Such a model is described by the Lagrangian
\be
\mathcal{L}_{0s0}=\frac{1}{2}(\partial_\mu \phi)^2-\frac{1}{2}m_\phi^2 \phi^2
-\frac{\lambda_\phi}{2\sqrt{2}}v h \phi^2
\label{L0s0}
\ee
with $v=246$ GeV. 
The DM coupling term of the form $\phi^4$ does not play a relevant role for LHC phenomenology and it will be neglected.

The low-energy Lagrangian (\ref{L0s0}) needs to be completed, at energies larger than $m_h$, in a gauge-invariant way, using the Higgs doublet $H$
\be
\mathcal{L}_{0s0}=\frac{1}{2}(\partial_\mu \phi)^2-\frac{1}{2}m_\phi^2 \phi^2
-\frac{\lambda_\phi}{4}  \phi^2 H^\dag H
\label{L0s0_low}
\ee
Note that this model is described by renormalizable interactions. A discrete $Z_2$  symmetry under which $H$ is even and $\phi$ is odd would make $\phi$ stable and prevent $\phi-H$ mixing.

The model parameters are simply $\{m_\phi, \lambda_\phi\}$ and one can distinguish two main regimes: $m_\phi<m_h/2$, $m_\phi>m_h/2$.

\paragraph{\textbf{\emph{Collider\\}}}

For DM lighter than half of the Higgs mass ($m_\phi<m_h/2$), the Higgs can decay on-shell to a DM pair. The main collider constraint comes from the invisible width of the Higgs, say $\Gamma_{h, inv}/\Gamma_h\lesssim 20\%$.
The Higgs to DM decay responsible for the invisible width is
\be
\Gamma(h\to \phi\phi)=\frac{\lambda_\phi^2}{32\pi}\frac{v^2}{m_h}
\sqrt{1-\frac{4m_\phi^2}{m_h^2}}.
\ee
Taking $\Gamma_h=4.2$ MeV for $m_h=125.6$ GeV, the 20\% constraint gives $\lambda_\phi\lesssim 10^{-2}$.

In the opposite regime ($m_\phi>m_h/2$), the invisible width constraint does not apply anymore. The cross-section for DM production at the LHC is further suppressed by $\lambda_\phi^2$ and phase space, thus making mono-jet  search strategies irrelevant. The most important constraint for this region of parameter space is on the spin-independent (SI) scattering cross-section from direct detection experiments.

\paragraph{\textbf{\emph{DM self-annihilation\\}}}

Using (\ref{L0s0}), the DM self-annihilation cross section to SM fermions of mass $m_f$ is
\be
\langle\sigma v_{\rm rel}\rangle(\phi\phi\to \bar f f)=
\frac{\lambda_\phi^2 m_f^2}{8\pi}
\frac{\left(1-\frac{m_f^2}{m_\phi^2}\right)^{3/2}}{(m_h^2-4m_\phi^2)^2+m_h^2\Gamma_h^2}+\mathcal{O}(v_{\rm rel}^2)
\ee
where $v_{\rm rel}$ is the relative velocity of DM particles.

Using the high-energy completion Eq.~(\ref{L0s0_low}),  the annihilation to $hh$ final states also opens up,
\be
\langle\sigma v_{\rm rel}\rangle(\phi\phi\to hh)=
\frac{\lambda_\phi^2}{512\pi m_\phi^2},
\qquad (m_h=0).
\ee

\paragraph{\textbf{\emph{DM scattering on nucleons\\}}}

The effective Lagrangian at the DM-quark level is
\be
\mathcal{L}_{\rm eff}=\sum_q \frac{y_q \lambda_\phi v}{4m_h^2} \phi^2 \bar q q
\ee
where the Yukawa coupling $y_q$ is defined by $m_q=y_q v/\sqrt{2}$. The DM-nucleon scattering cross section is given by Eq.~(\ref{sigSI}), with coefficient (cf. Table \ref{ScalarDMNucleonList})

\bea
c^N&=&\sum_{q=u,d,s} \frac{y_q \lambda_\phi v}{4m_h^2} \frac{m_N}{m_q} f^{(N)}_{q} + \frac{2}{27}f^{(N)}_{G}\left( \sum_{q=c,b,t} \frac{y_q \lambda_\phi v}{4m_h^2} \frac{m_N}{m_q}\right)\nonumber\\
&=& \frac{ \lambda_\phi m_N}{2\sqrt{2} m_h^2} \left(\sum_{q=u,d,s}  f^{(N)}_{q} + \frac{6}{27}\left(1- \sum_{q=u,d,s} f^{(N)}_{q}\right)\right),
\eea
where $f^{(N)}_{q}$ are given in Table~\ref{DDConstants}, and recalling that $f^{(N)}_{G}\equiv1- \sum_{q=u,d,s} f^{(N)}_{q}$

\subsubsection{Fermion DM, $s$-channel ($0s\frac{1}{2}$ model)}
\label{subsec:0s12}

\paragraph{\textsf{\emph{$\blacksquare$ GENERIC CASE\\}}}

The next case we would like to consider is a spin-1/2 DM particle, taken to be a Dirac fermion. The Majorana case only involves some minor straightforward changes.

We consider two benchmark models where the gauge-singlet mediator is either a scalar $S$ or a pseudoscalar $A$,  described by the Lagrangians \cite{1410.6497,1511.06451}

\bea
\mathcal{L}_{0_S s\frac{1}{2}}&=&\frac{1}{2} (\partial_\mu S)^2-\frac{1}{2}m_S^2 S^2+\bar\chi(i\slashed{\partial}-m_\chi)\chi
-g_\chi S\bar\chi\chi- g_{\rm SM}S\sum_f \frac{y_f}{\sqrt{2}}\bar f f\,,
\label{0Ss12}
\\
\mathcal{L}_{0_A s\frac{1}{2}}&=&\frac{1}{2} (\partial_\mu A)^2-\frac{1}{2}m_A^2 A^2+\bar\chi(i\slashed{\partial}-m_\chi)\chi
-ig_\chi A\bar\chi\gamma^5\chi- i g_{\rm SM}A\sum_f \frac{y_f}{\sqrt{2}}\bar f\gamma^5 f\,,
\label{0As12}
\eea
where the sum runs over SM fermions $f$.
The DM particle is unlikely to receive its mass from electroweak symmetry breaking, so its interaction with the mediator has not been set proportional to a Yukawa coupling.

As for the operators connecting the mediators to SM fermions, the MFV hypothesis requires the coupling to be proportional to the Yukawas $y_f$. However, in full generality it is possible to have non-universal $g_{\rm SM}$ couplings, e.g. $g_{\rm SM}^{(u)}\neq g_{\rm SM}^{(d)}\neq g_{\rm SM}^{(\ell)}$, for up-type quarks,  down-type quarks and  leptons. Notably, the situation where $g_{\rm SM}^{(u)}\neq g_{\rm SM}^{(d)}$
arises in Two-Higgs-Doublet Models. In the following we will focus on the universal couplings, but the reader should keep in mind that this is not the most general situation.

Another caveat concerns the mixing of the scalar mediator with the Higgs. In general, Lagrangian operators mixing a gauge singlet scalar with a Higgs doublet (e.g. $S^2 |H|^2$) are allowed. As discussed in the introduction of Section \ref{subsec:scalarmediator} we will follow the common assumption by neglecting these mixings. However, we will later discuss an example of these when discussing the Scalar-Higgs Portal.

So, the simplified models described by equations (\ref{0Ss12}) and (\ref{0As12})  have a minimal parameter count:
\be
\{m_\chi, m_{S/A}, g_\chi, g_{\rm SM}\}\,.
\ee

\paragraph{\textbf{\emph{Collider\\}}}

The mediators have decay channels to SM fermions, DM particles or to gluons (via a fermion loop dominated by the top-quark). The corresponding partial widths are
\bea
\Gamma(S/A\to \bar f f)&=&\sum_f N_c(f)\frac{y_f^2 g_{\rm SM}^2 m_{S/A}}{16\pi}
\left(1-\frac{4m_f^2}{m_{S/A}^2}\right)^{n/2}\\
\Gamma(S/A\to \bar \chi \chi)&=&\frac{ g_{\chi}^2 m_{S/A}}{8\pi}
\left(1-\frac{4m_\chi^2}{m_{S/A}^2}\right)^{n/2}\\
\Gamma(S/A\to gg)&=&\frac{\alpha_s^2 g_{\rm SM}^2 }{32\pi^3}
\frac{m_{S/A}^3}{v^2}\left| f_{S/A}\left(\frac{4m_t^2}{m_{S/A}^2}\right)\right|^{2}
\eea
where $N_c(f)$ is the number of colors of fermion $f$ (3 for quarks, 1 for leptons), and $n=1$ for pseudoscalars,  3 for scalars. The loop functions are
\bea
f_S(\tau)&=&\tau\left[1+(1-\tau)\arctan^2\frac{1}{\sqrt{\tau-1}}\right]\\
f_A(\tau)&=&\tau\arctan^2\frac{1}{\sqrt{\tau-1}}
\eea
for $\tau>1$.

Other loop-induced decay channels, such as decay to $\gamma\gamma$, are sub-dominant. Of course, in the presence of additional (possibly invisible) decay modes of the mediators, the total width will be larger than the sum of the partial widths written above.

Typically, the decay to DM particles dominates, unless the mediator is heavy enough to kinematically open  the decay to top-quarks. Also notice  the different scaling with respect to DM velocity $(1-4m_f^2/m_{S/A}^2)^{n/2}$ for scalars and pseudoscalars. In the region close to the kinematic boundary, the decay width of $A$ is larger and therefore one expects stronger constraints on pseudoscalars than on scalars.

\begin{figure}[t!]
\centering
\includegraphics[width=0.3\textwidth]{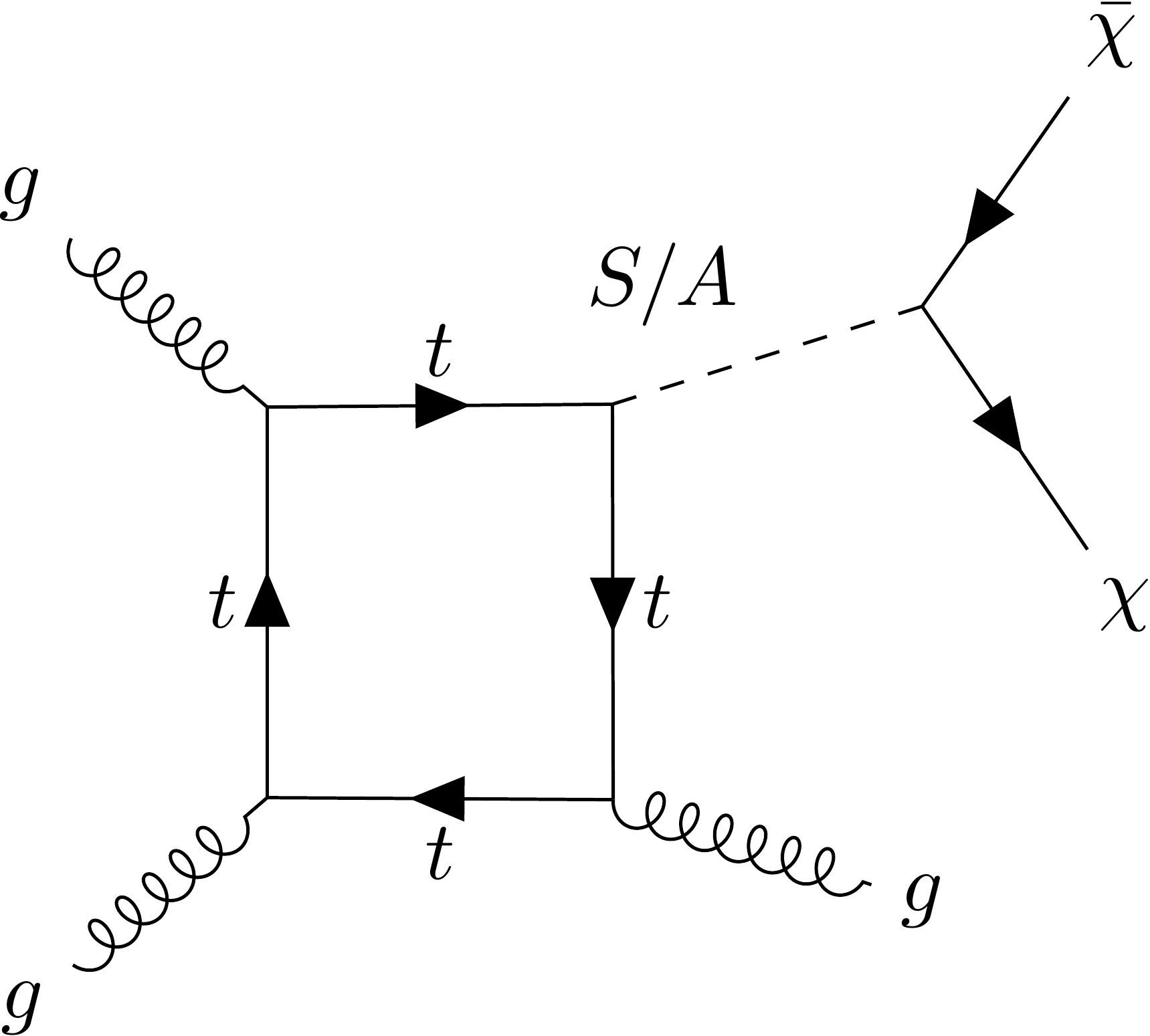}\quad
\includegraphics[width=0.3\textwidth]{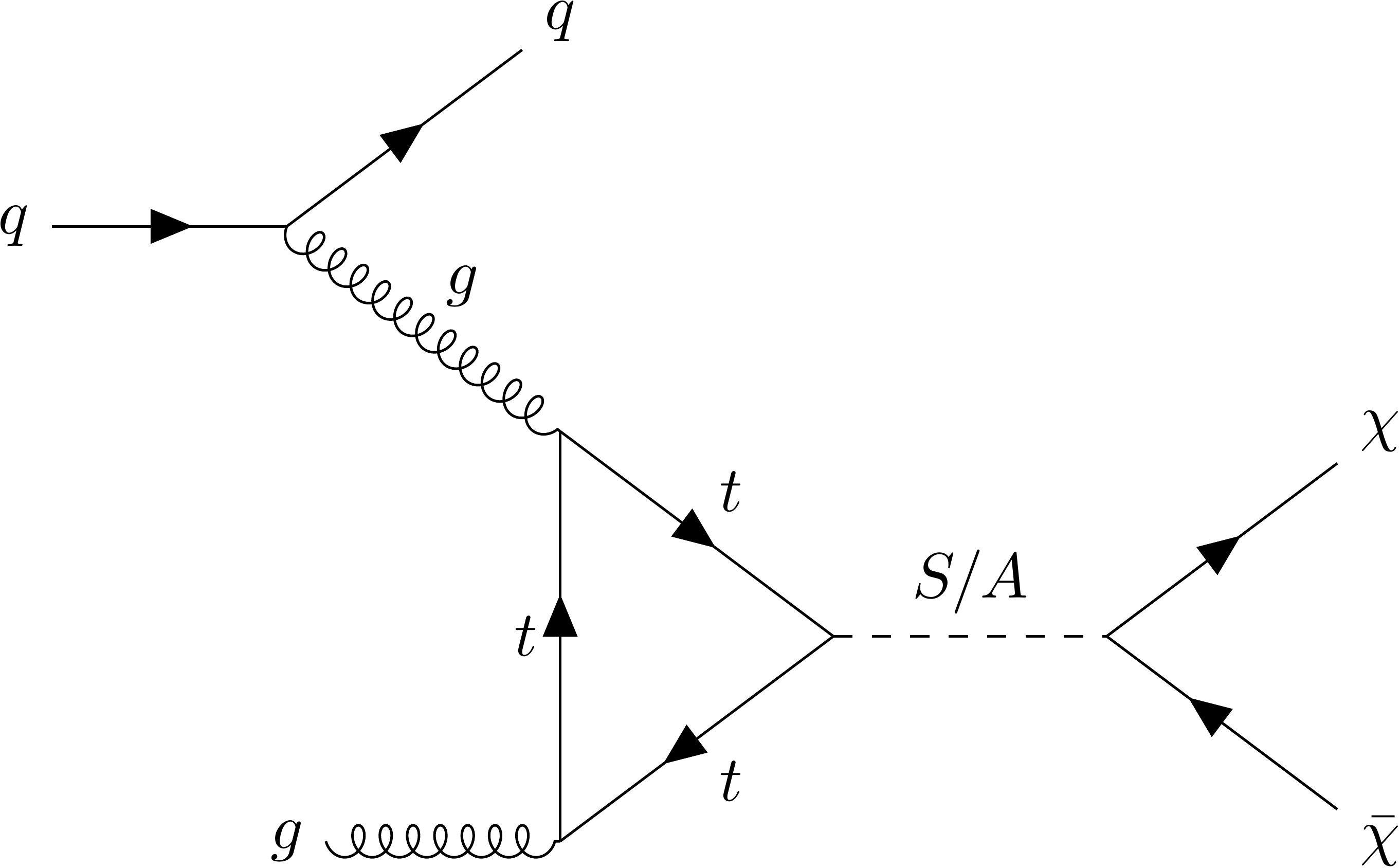}\quad
\includegraphics[width=0.3\textwidth]{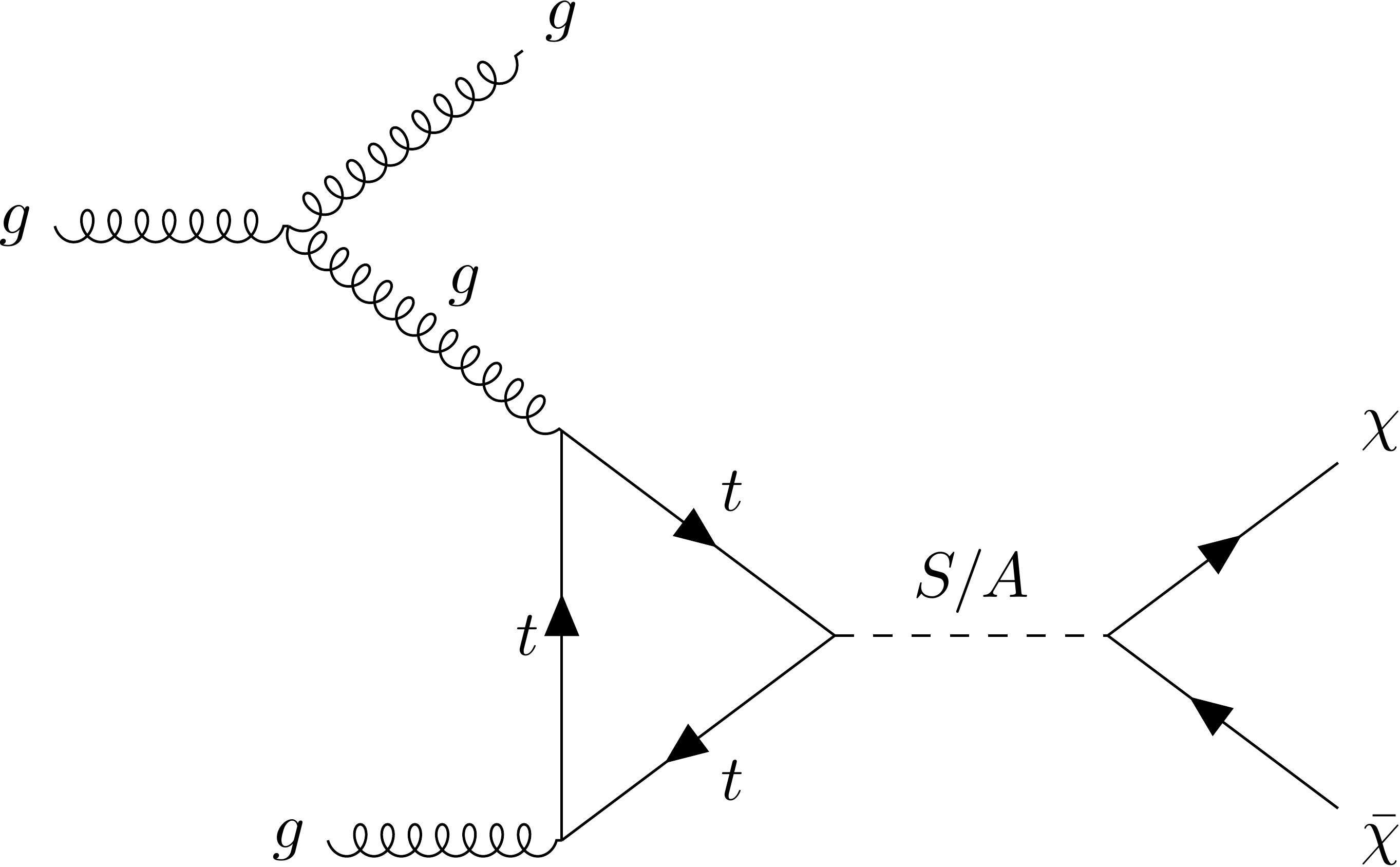}\\
\vspace{0.2cm}
\includegraphics[width=0.35\textwidth]{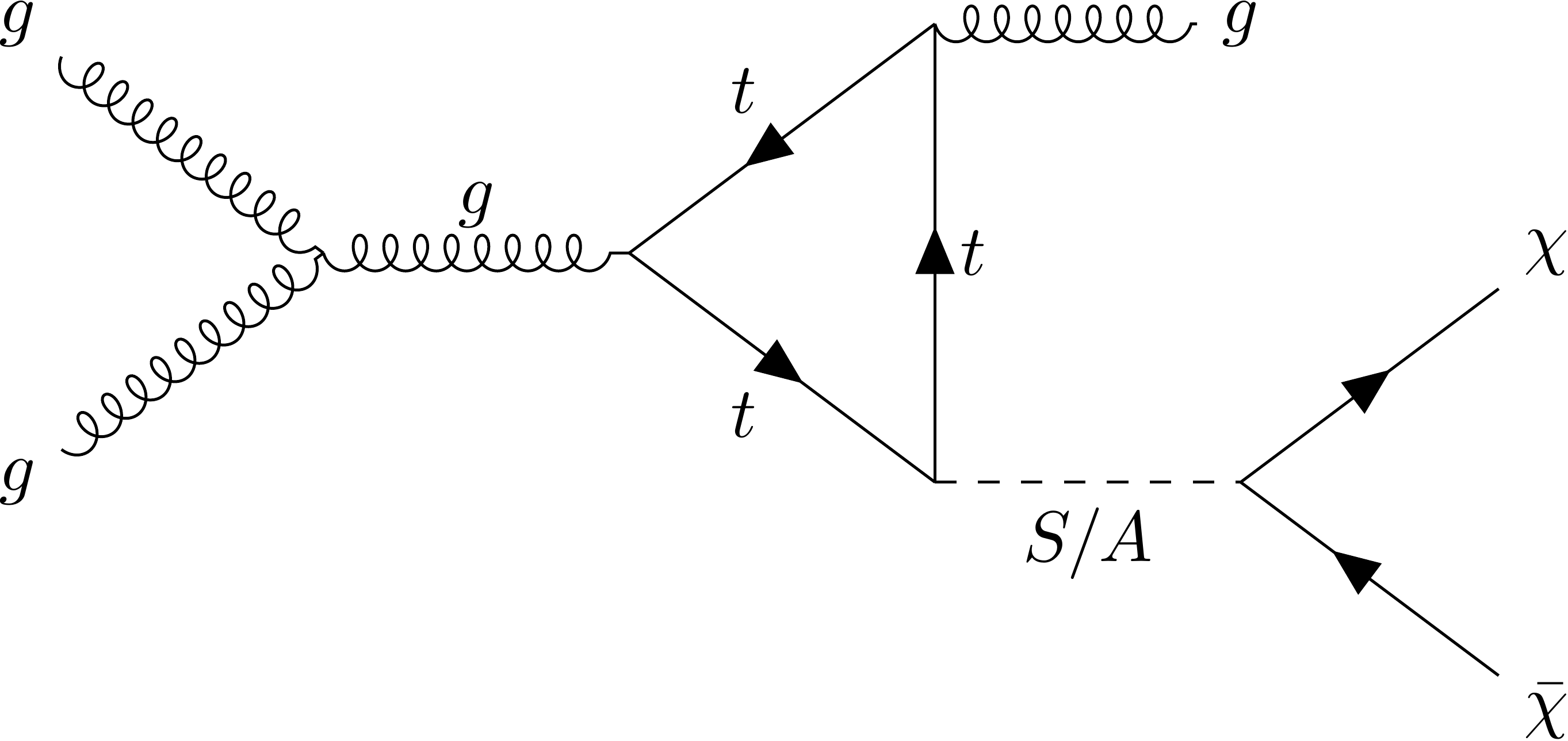}\qquad
\includegraphics[width=0.2\textwidth]{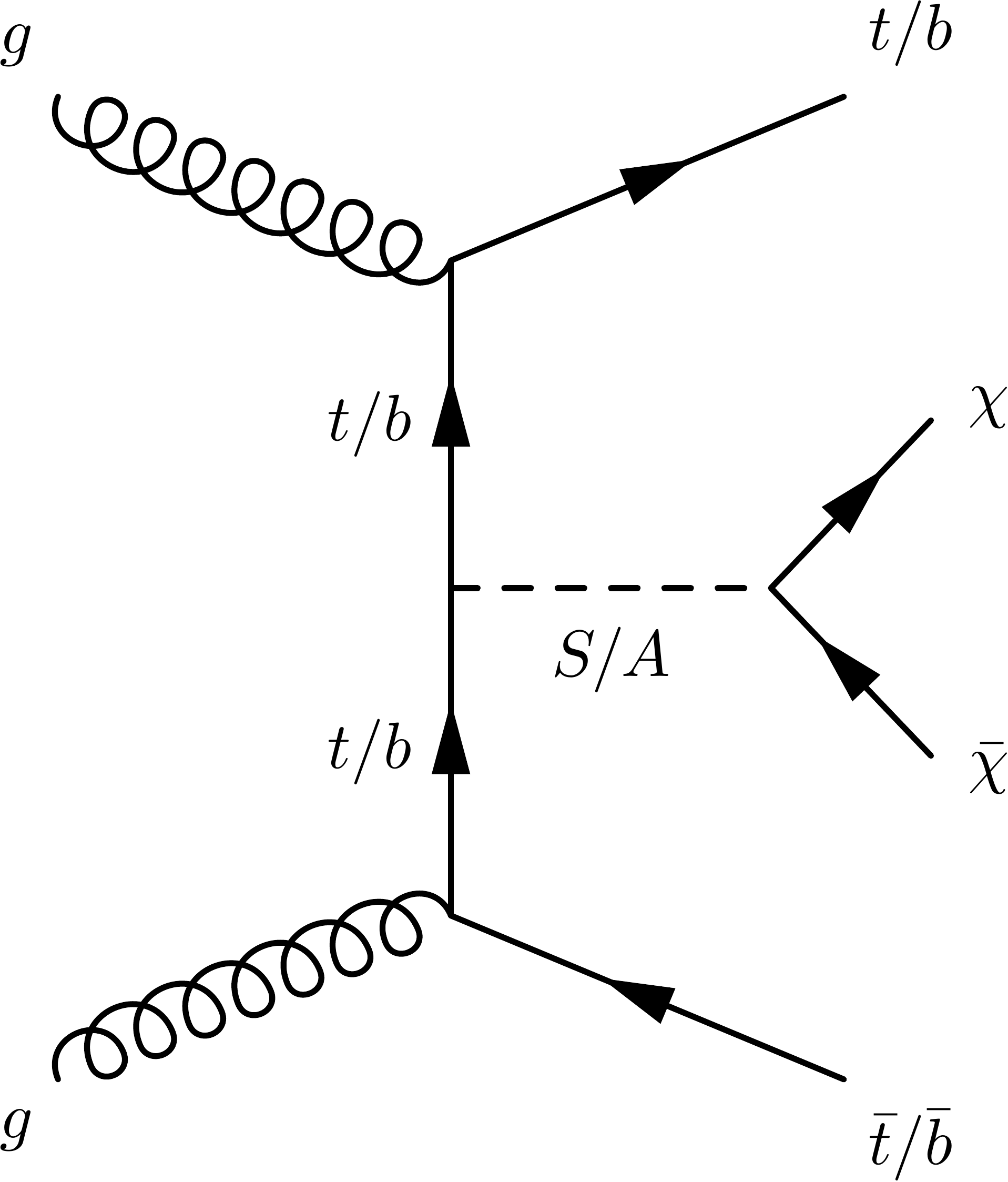}
\caption{\em 
Diagrams contributing to $\missET+j$, $\missET+t\bar t$ and $\missET+b\bar b$  signals.
The $\missET+j$ diagrams involve loop of top-quarks. while $\missET+t\bar t, \missET+b\bar b$ involve tree-level emission of mediator from a t-channel top-quark exchange. Most Feynman diagrams were generated using Ti$k$Z-Feynman \cite{1601.05437}.}
\label{fig:0s12collider}
\end{figure}

There are three main  strategies to search for this kind of simplified model at colliders:
missing energy (MET) with 1 jet  ($\missET+j$), MET with 2 top-quarks ($\missET+t\bar t$), MET with 2 bottom-quarks ($\missET+b\bar b$), see Fig.~\ref{fig:0s12collider}. 
Much recent and ongoing effort has gone into improving predictions for these signals by including next-to-leading-order (NLO) QCD effects in simulations of the signals from these and other simplified models  \cite{1310.4491,1508.00564,1508.05327,1509.05785}.

The $\missET+j$ searches are expected to provide the strongest discovery potential, but the channels with heavy quarks tagged can have much lower backgrounds, and they can get more and more relevant as the energy and the luminosity of LHC is increasing.

\paragraph{\textbf{\emph{DM self-annihilation\\}}}

The self-annihilations of two DM particles are the key processes to consider when studying the relic abundance (freeze-out mechanism in the early universe) or the indirect detection constraints (constraints from observations of DM annihilation products, usually studying annihilation in the halo or galactic center today). 

The thermally averaged self-annihilation cross sections of Dirac DM $\chi$, via a scalar or pseudoscalar mediator, to SM fermions $f$ are
\bea
\langle\sigma v_{\rm rel}\rangle(\phi\phi\to S\to \bar f f)&=&N_c(f)
\frac{g_\chi^2 g_{\rm SM}^2 y_f^2}{16\pi}
\frac{m_\chi^2\left(1-\frac{m_f^2}{m_\chi^2}\right)^{3/2}}{(m_S^2-4m_\chi^2)^2+m_S^2\Gamma_S^2}v_{\rm rel}^2 \label{sigv0s12}\\
\langle\sigma v_{\rm rel}\rangle(\phi\phi\to A\to \bar f f)&=&N_c(f)
\frac{g_\chi^2 g_{\rm SM}^2 y_f^2}{4\pi}
\frac{m_\chi^2\left(1-\frac{m_f^2}{m_\chi^2}\right)^{1/2}}{(m_A^2-4m_\chi^2)^2+m_A^2\Gamma_A^2}\,.
\eea
For Majorana DM, the above cross-sections get multiplied by 2. Notice that the annihilation via scalar mediator is in $p$-wave ($v^2$-suppressed) even for $m_f\neq 0$.

\paragraph{\textbf{\emph{DM scattering on nucleons\\}}}

In the low-energy regime at which DM-nucleon scattering is taking place, it is possible to integrate out the mediator and recover the EFT description,  with the operators 
\bea
\mathcal{O}_S=\frac{g_\chi g_{\rm SM}y_q}{\sqrt{2}m_S^2}(\bar \chi \chi)(\bar q q)
=\frac{g_\chi g_{\rm SM}y_q}{\sqrt{2}m_S^2}\Op{D1}\\
\mathcal{O}_A=\frac{g_\chi g_{\rm SM}y_q}{\sqrt{2}m_A^2}(\bar \chi i\gamma^5 \chi)(\bar q i\gamma^5 q)
=\frac{g_\chi g_{\rm SM}y_q}{\sqrt{2}m_A^2}\Op{D4}
\eea
 describing the DM-quarks fundamental scattering, and expressed in terms of the operators in Table \ref{DList}. Remember that the operator coefficients must be evaluated at the scale where scattering is occurring \cite{1402.1173,1411.3342},  by performing RG evolution from the high energy theory as well as matching conditions at the quark mass thresholds.
 
 The scalar exchange gives rise to spin-independent DM-nucleon scattering, while the pseudoscalar gives a spin and momentum suppressed cross-section. The latter case does not provide significant constraints from direct detection experiments. As for the SI case, the elastic DM-nucleon cross section (for Dirac DM) is given by Eq.~(\ref{sigSI}),
with effective coupling (cf. Table \ref{DMNucleonList})
\bea
c^N&=&\sum_{q=u,d,s} f_q^{(N)}\frac{m_N}{m_q} \left(\frac{g_\chi g_{\rm SM}y_q}{\sqrt{2}m_S^2}\right)
+\frac{2}{27}f_G^{(N)}\sum_{q=c,b,t} \frac{m_N}{m_q} \left(\frac{g_\chi g_{\rm SM}y_q}{\sqrt{2}m_S^2}\right)\nn\\
&=&
\left(\frac{g_\chi g_{\rm SM}}{ v m_S^2}\right)m_N\left[
\sum_{q=u,d,s} f_q^{(N)}+\frac{6}{27}\left(1-\sum_{q=u,d,s} f_q^{(N)}\right)
\right]\label{sigSI0s12}
\,.
\eea
where we have again used that $f_G^{(N)}=1-\sum_{q=u,d,s}f_q^{(N)}$ and
that $g_{\rm SM}$ was assumed to be flavor-universal, otherwise one cannot take that factor out of the sum over quarks.
Sample numerical values of the couplings $f_q^{(N)}$ are listed in Table \ref{DDConstants}.

\paragraph{\textsf{\emph{$\blacksquare$ CASE STUDY 1: HIGGS AS MEDIATOR\\}}}

As a first case study, we consider one specific realisation of the $0s\frac{1}{2}$ model outlined earlier, where the Higgs itself serves as the scalar mediator particle. We already considered this possibility in Section \ref{subsec:0s0} for the case of scalar DM ($0s0$ model), and we will consider another scenario involving both Higgs and vector mediators in Case Study 4 on Page \pageref{casestudy4}. Here we want to outline the main features of this ``Higgs portal'' model for Dirac fermion DM  \cite{0803.2932,1112.1847,1203.2064,1203.4854,1312.2592,1408.4929,1504.03610}.

The Lagrangian  of the model at low energies is
\be
\mathcal{L}\supset-\frac{h}{\sqrt{2}}\left[
\sum_fy_f\bar f f+\bar \chi(y_\chi+iy_\chi^P \gamma^5)\chi
\right]
\label{HiggsPortalL}
\ee
 which can be matched to the Lagrangians Eqs.~(\ref{0Ss12}) and (\ref{0As12}) of Section \ref{subsec:0s12},   provided that $y_\chi=g_\chi\sqrt{2}$, or $y_\chi^P=g_\chi\sqrt{2}$, $g_{\rm SM}=1$.

Notice, however, that here the Higgs $h$ is a real scalar field (not a pseudoscalar, like the generic mediator $A$); so the pseudoscalar coupling in Eq.~(\ref{HiggsPortalL}) only affects the $h$-DM interaction and not the usual Yukawa interactions between the Higgs and the SM  fermions $f$. So the generic pseudoscalar model $0_As\frac{1}{2}$ cannot be completely matched with the model in Eq.~(\ref{HiggsPortalL})  since the Higgs is a real scalar.
 
At energies larger than the Higgs mass, the effective Lagrangian in Eq.~(\ref{HiggsPortalL}) is completed in a gauge-invariant way as
\be
\mathcal{L}\supset-\frac{H^\dag H}{2 v}\bar \chi(y_\chi+iy_\chi^P \gamma^5)\chi
\label{HiggsPortalLhighenergy}
\ee
which is described by a dimension-5 operator.

The model parameters are simply $\{m_\chi, y_\chi\}$ or  $\{m_\chi, y_\chi^P\}$,  if one considers the scalar and pseudoscalar couplings separately.
 
\paragraph{\textbf{\emph{Collider\\}}}

For DM lighter than half of the Higgs mass ($m_\chi<m_h/2$), the on-shell decays of the  Higgs into a DM pair contribute to the Higgs invisible width
\be
\Gamma(h\to \bar\chi\chi)=\frac{y_\chi^2 m_h}{16\pi}\left(
1-\frac{4m_\chi^2}{m_h^2}
\right)^{3/2}
\ee
or
\be
\Gamma(h\to \bar\chi\chi)=\frac{(y_\chi^P)^2 m_h}{16\pi}\left(
1-\frac{4m_\chi^2}{m_h^2}
\right)^{1/2}
\ee
The experimental constraint $\Gamma_{h,inv}/\Gamma_h\lesssim 20\%$ gives $y_\chi, y_\chi^P\lesssim 10^{-2}$, for $\Gamma_h=4.2$ MeV and $m_h=125.6$ GeV\cite{1402.6287}.

The opposite mass regime $m_\chi>m_h/2$ is not significantly constrained by collider data, for couplings within the perturbative domain.

\paragraph{\textbf{\emph{DM self-annihilation\\}}}

The thermally-averaged annihilation cross sections for Dirac fermion DM are
\bea
\langle\sigma v_{\rm rel}\rangle(\chi\chi\to \bar f f)&=&N_c(f)
\frac{y_\chi^2y_f^2}{32\pi}
\frac{m_\chi^2\left(1-\frac{m_f^2}{m_\chi^2}\right)^{3/2}}{(m_h^2-4m_\chi^2)^2+m_h^2\Gamma_h^2}v_{\rm rel}^2\\
\langle\sigma v_{\rm rel}\rangle(\chi\chi\to \bar f f)&=&N_c(f)
\frac{(y_\chi^P)^2y_f^2}{8\pi}
\frac{m_\chi^2\left(1-\frac{m_f^2}{m_\chi^2}\right)^{3/2}}{(m_h^2-4m_\chi^2)^2+m_h^2\Gamma_h^2}\,.
\eea 
For Majorana DM one needs to include an extra factor of 2. 

The scalar coupling does not produce $s$-wave cross sections. For DM masses above the Higgs mass, the Lagrangian operator Eq.~(\ref{HiggsPortalLhighenergy})  opens up self-annihilations to two Higgses or longitudinal gauge bosons
\be
\langle\sigma v_{\rm rel}\rangle(\chi\chi\to  H H)=
\frac{1}{64\pi v^2}
\left[
(y_\chi^P)^2+\frac{v_{\rm rel}^2}{4}y_\chi^2
\right]
\,.
\ee

\paragraph{\textbf{\emph{DM scattering on nucleons\\}}}

At low energies, after integrating out the Higgs field, we end up with the effective Lagrangian
\be
\mathcal{L}_{\rm eff}\supset \frac{y_f}{2m_h^2}(\bar q q)\left[
\bar\chi(y_\chi+i\gamma^5y_\chi^P)\chi
\right]\,.
\ee
The coupling $y_\chi$ multiplies the $\mathcal{O}_{D1}$ operator while $y_\chi^P$ is in front of a $\mathcal{O}_{D2}$ operator. Therefore, the scalar coupling is responsible for spin-independent cross section, while the pseudoscalar coupling drives a spin and momentum dependent cross-section, as described in section~\ref{sec:eft-dd}. The spin-independent cross section can be found via Eqn.~(\ref{sigSI}) with coefficient
\be
c^N=\frac{y_\chi m_N}{\sqrt{2}v m_h^2}
\left[\sum_{q=u,d,s}f_q^{(N)}+\frac{6}{27}(1-\sum_{q=u,d,s}f_q^{(N)})\right]
\ee
The current best limits on spin-independent cross-section from LUX \cite{1512.03506} rule out a fermion DM coupling to Higgs with the correct thermal relic abundance for  $m_\chi\lesssim 10^3$ GeV. However, unknown particles/interactions may reduce the abundance of DM coupled to Higgs and relax the tension with DD data. 

On the other hand, because of much weaker constraints on spin and momentum suppressed cross sections, there are currently no limits on perturbative values of $y_\chi^P$ from direct detection, thus  leaving this case as still viable.

\paragraph{\textsf{\emph{$\blacksquare$ CASE STUDY 2: SCALAR-HIGGS PORTAL\\}}}

Another specific realization of the $0s\frac{1}{2}$ model arises by allowing mixing between a real scalar mediator $S$ and the Higgs boson. In this case, to keep the model as minimal as possible, the mediator $S$ is not allowed to have couplings directly to the SM fermions, but only through the ``Higgs portal''. 
Therefore, this kind of model looks like a Two-Higgs-Doublet Model (2HDM) extension of the SM Higgs sector (see Ref.~\cite{1106.0034} for a review and Refs.~\cite{1404.3716,1509.01110} for some recent work on the pseudoscalar mediator case). 
The DM is again assumed to be a Dirac fermion and the Lagrangian describing the model is
\bea
\mathcal{L}&\supset&\frac{1}{2}(\partial_\mu S)^2-\frac{1}{2}m_S^2 S^2
+\bar\chi(i\slashed{\partial}-m_\chi)\chi 
- \frac{h}{\sqrt{2}}\sum_f y_f \bar f f\nn\\
&&-y_\chi S\bar \chi\chi-\mu_S S|H|^2-\lambda_S S^2 |H|^2
\,.
\label{2HDMLagr}
\eea
The cubic and quartic self-couplings of the mediator $S$ do not play any role for LHC phenomenology and they have not been considered in the Lagrangian. 
Another simplification is to forbid the $S$ mediator from developing a VEV, $\langle S\rangle=0$. 
The generalization where this assumption is relaxed 
is straightforward. 

This model is described by the 4 parameters: $\{m_\chi, m_S, \lambda_S, \mu_S\}$. 
The mediator-Higgs mixing driven by $\mu_S$ leads us to diagonalize the mass matrix and find the physical mass eigenstates $h_1$ and $h_2$
\be
\left(
\begin{matrix}
h_1\\ h_2 
\end{matrix}
\right)
=
\left(
\begin{matrix}
\cos\theta&\sin\theta\\ 
-\sin\theta &\cos\theta
\end{matrix}
\right)
\left(
\begin{matrix}
h\\ S
\end{matrix}
\right)\,,
\ee
where the mixing angle is defined by $\tan(2\theta)=2 v\mu_S/(m_S^2-m_h^2+\lambda_S v^2)$, in such a way that $\theta=0$ ($\mu_S=0$) corresponds to a dark sector decoupled from the SM, and the physical masses are approximately given by
\bea
m_{h_1}&\simeq& m_h\\
m_{h_2}&\simeq& \sqrt{m_S^2+\lambda_S^2 v^2}\,,
\eea
so that $h_1$ corresponds to the physical Higgs boson of mass $\sim 125$ GeV.

In the mass-eigenstate basis, the Lagrangian (\ref{2HDMLagr}) reads
\be
\mathcal{L}\supset 
-(h_1 \cos\theta-h_2\sin\theta)\sum_f \frac{y_f}{\sqrt{2}}\bar f f
-(h_1\sin\theta+h_2\cos\theta)y_\chi \bar \chi \chi\,.
\ee
This Lagrangian is of the same form as the generic one
$\mathcal{L}_{0_Ss\frac{1}{2}}$ of Eq.~(\ref{0Ss12}), where
we can identify $h_2$ with $S$ and read the corresponding
couplings
\bea
g_\chi&=&y_\chi\cos\theta\label{2HDMa}\\
g_{\rm SM}&=&-\sin\theta\,,\label{2HDMb}
\eea
while the Higgs Yukawa couplings to fermions are reduced as $y_f\cos\theta$.

\paragraph{\textbf{\emph{Collider\\}}}

In addition to the Yukawa couplings, the $\cos\theta$ suppression also appears  in the trilinear couplings of the Higgs with two gauge bosons, and therefore $\theta$ is constrained by Higgs physics measurements as well as EW precision tests. The limits from LHC Run I Higgs physics are the most stringent ones and give $\sin\theta\lesssim 0.4$
\cite{1302.5694, 1308.0295}.

The invisible width of the Higgs decaying to DM particles is
\be
\Gamma(h_1\to \bar \chi \chi)\frac{y_\chi^2 \sin^2\theta m_{h_1}}{8\pi}\left(1-\frac{4m_\chi^2}{m_{h_1}^2}\right)^{3/2}\,,
\ee
and for a light enough mediator, the $h_1\to h_2 h_2$ decay can  also open up. The calculation of the invisible BR of the Higgs should also take into account that the Higgs decays to SM fermions receive a $\cos^2\theta$ suppression.

\begin{figure}[t!]
\centering
\includegraphics[width=0.45\textwidth]{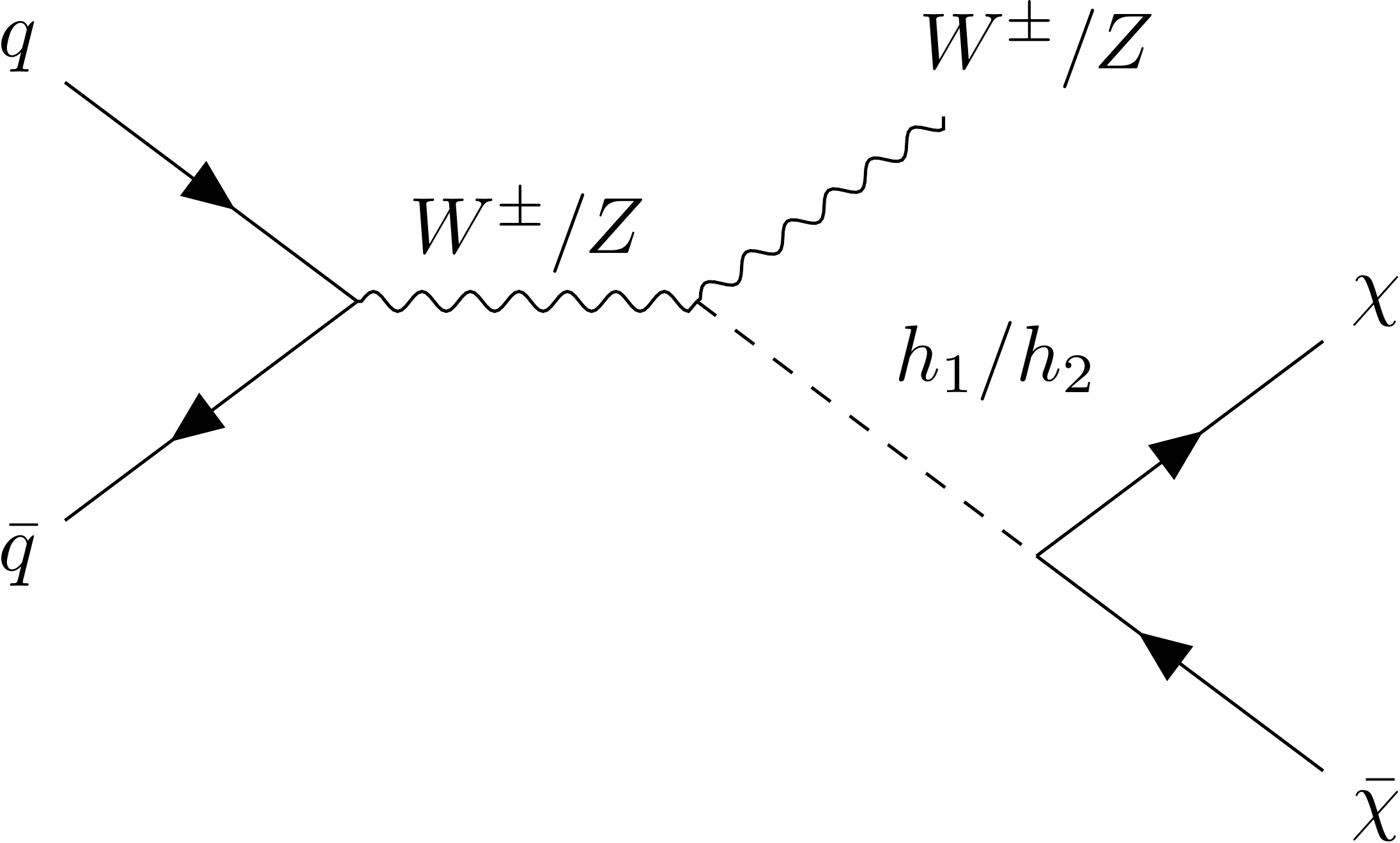}
\caption{\em 
Diagram of the process contributing to mono-W/Z signals in Scalar-Higgs Portal.}
\label{fig:2HDM}
\end{figure}

On top of the usual $\missET+j$ signal, this 2HDM-like simplified model possesses other interesting channels that
may distinguish it from the generic scalar mediator case.
For instance, mono-W/Z signals can arise at tree level as in 
Fig.~\ref{fig:2HDM}.

An important feature is the destructive interference between the exchange of $h_1$ and $h_2$, which has an impact on both  LHC and DD phenomenology.

\begin{figure}[t!]
\centering
\includegraphics[width=0.4\textwidth]{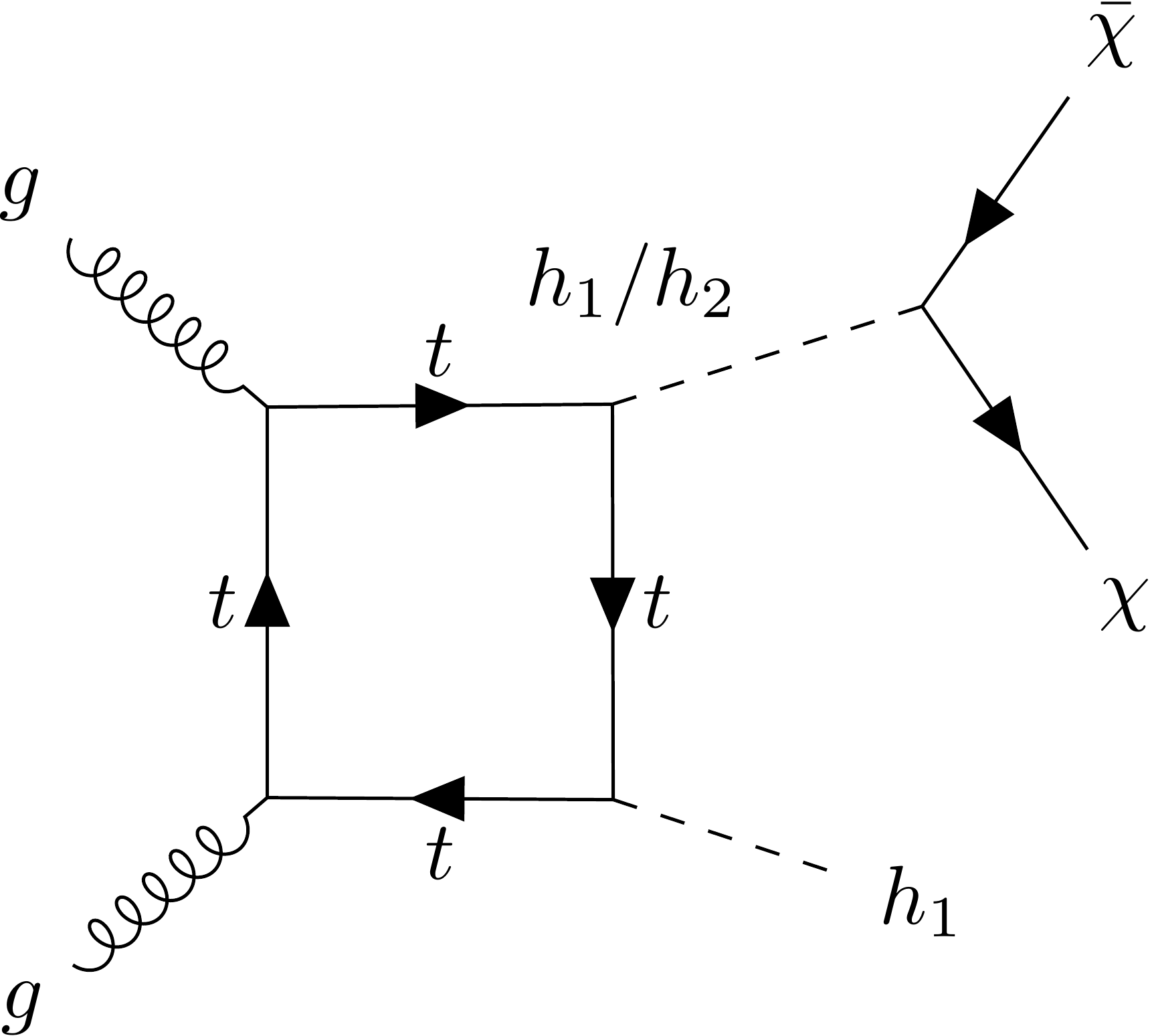}
\qquad
\includegraphics[width=0.4\textwidth]{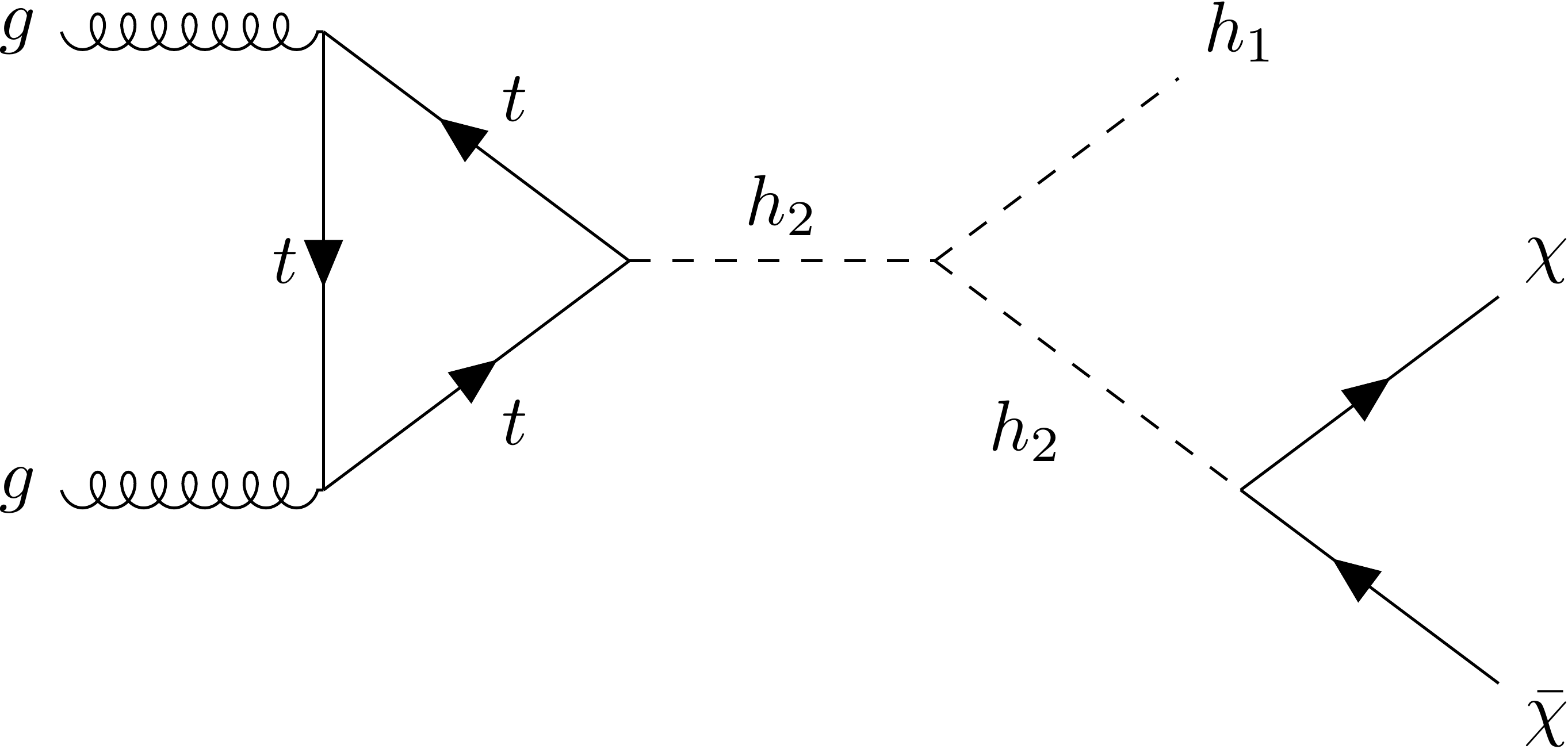}
\caption{\em 
Diagrams contributing to mono-Higgs signals in Scalar-Higgs Portal.}
\label{fig:2HDM2}
\end{figure}

Furthermore, the $h_1 h_2^2$ trilinear vertex is likely to change the phenomenology of mono-Higgs signals by adding to the usual diagram (Fig.~\ref{fig:2HDM2} left), and the diagram with triangle top-loop (Fig.~\ref{fig:2HDM2} right).

\paragraph{\textbf{\emph{DM self-annihilation and scattering on nucleons\\}}}

The DM self-annihilation rate and scattering rate are identical to the generic case with scalar mediator described by Eqs.~(\ref{sigv0s12}), (\ref{sigSI0s12}) respectively, with the couplings $g_\chi,g_{\rm SM}$ replaced with the expressions in Eqs.~(\ref{2HDMa}), (\ref{2HDMb}).

\subsubsection{Fermion DM, $t$-channel ($0t\frac{1}{2}$ model)}
\label{subsec:0t12}

Let us now turn to consider the most common situation of this kind, where the DM is a spin-$\frac{1}{2}$ (Dirac or Majorana) fermion $\chi$ and the mediator is a scalar particle $\eta$.
The interaction of interest is the one connecting $\chi$ and $\eta$ to a quark field $q$: $\eta \bar\chi q+$h.c. .
Since DM cannot have color charge, $\eta$ has to be colored. As for flavor, in order to comply with MFV, either $\eta$ or $\chi$ should carry a flavor index.
Although models with flavored DM has been considered
\cite{1109.3516, 1308.0584, 1404.1373}, we consider here the situation of unflavored DM where $\eta$ carries flavor index \cite{1308.2679, 1503.01500,  1403.4634, 1104.3823, 1512.00476,1307.8120, 1308.0612, 1209.0231, 1308.0592, 1308.2679, 1402.2285, 1405.3101}.
In this case the mediator closely resembles the squarks of the MSSM, for which extensive searches already exist 
(see e.g. \cite{1105.2838}).

Having decided that $\eta$ carries both color and flavor indices, it remains to be seen whether it couples to right-handed quark singlets (up-type or down-type) or to left-handed quark doublets.
The choice made here is to couple $\eta$ to right-handed up-type quarks $u_i=\{u_R, c_R, t_R\}$, so that the Lagrangian for the 3 mediator species $\eta_i$ reads
\be
\mathcal{L}_{0t\frac{1}{2}}\supset
\sum_{i=1,2,3}\left[
\frac12(\partial_\mu\eta_i)^2-\frac12M_i^2 \eta_i^2
+(g_i\eta_i^* \bar\chi u_i+\textrm{ h.c.})
\right]\,.
\label{0t12lagr}
\ee
Other choices for mediator-quark interactions can be worked out similarly.

The MFV hypothesis imposes universal masses and couplings $M_1=M_2=M_3\equiv M$ and $g_1=g_2=g_3\equiv g$, thus resulting in a three-dimensional parameter space
\be
\{m_\chi, M, g\}\,.
\ee
However, the breaking of this universality is possible, resulting in a splitting of the third-generation mediator ($i=3$) from the first two ($i=1,2$).

Stability of DM against decays is ensured by considering $m_\chi<m_\eta$, so that DM decays are not kinematically open.

\paragraph{\textbf{\emph{Collider\\}}}

\begin{figure}[t!]
\centering
\includegraphics[width=0.3\textwidth]{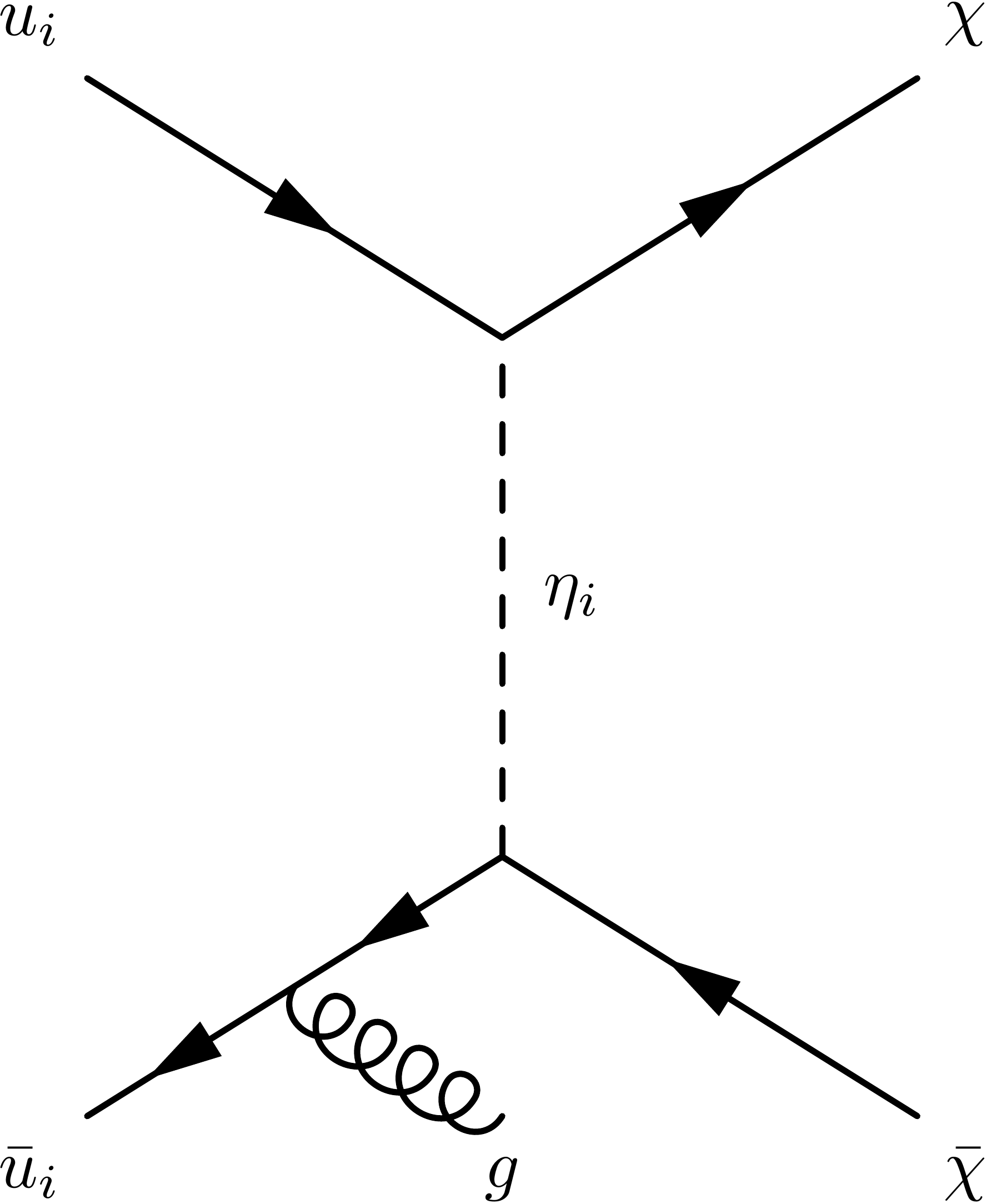}\qquad
\includegraphics[width=0.3\textwidth]{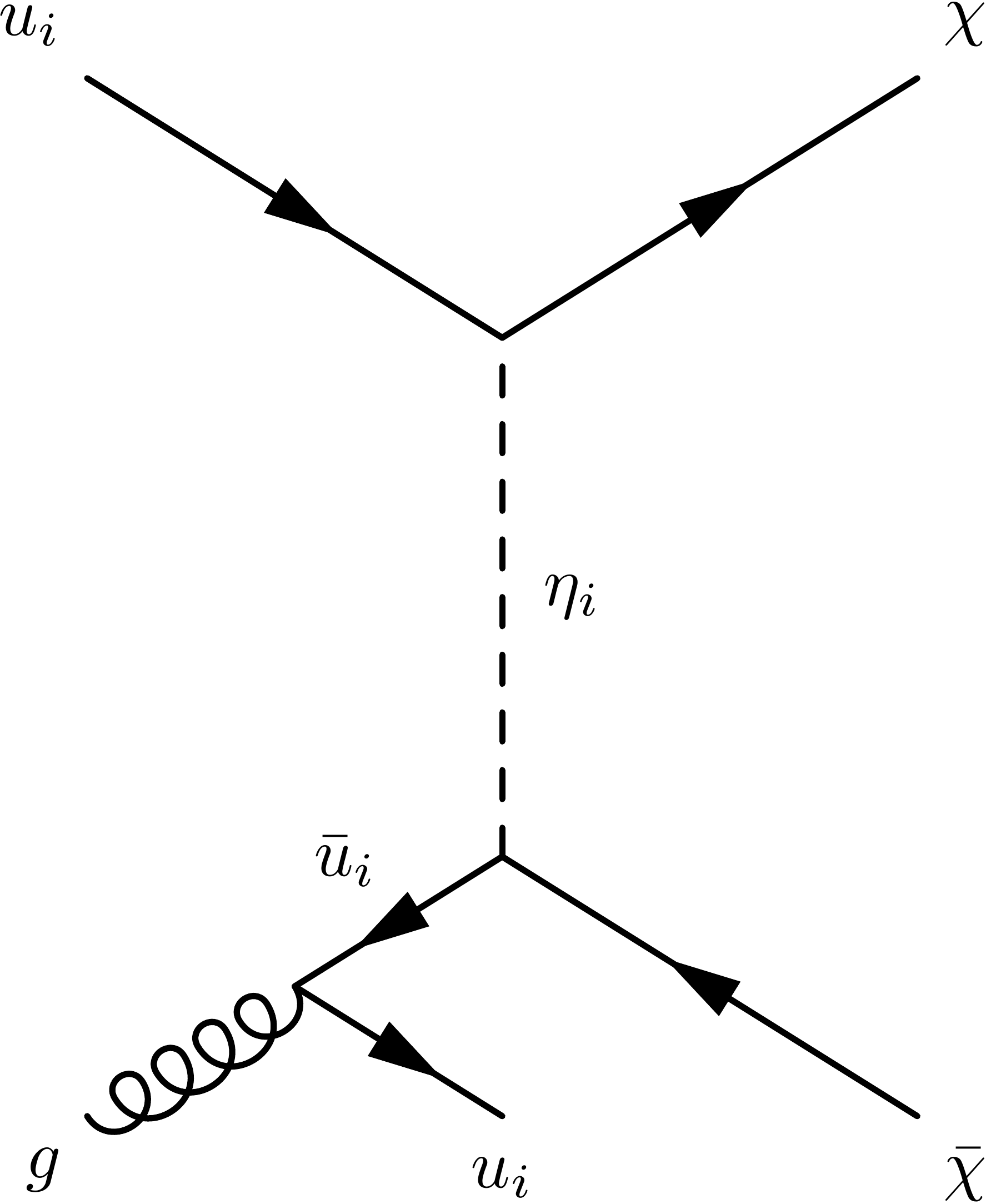}\\
\vspace{0.2cm}
\includegraphics[width=0.4\textwidth]{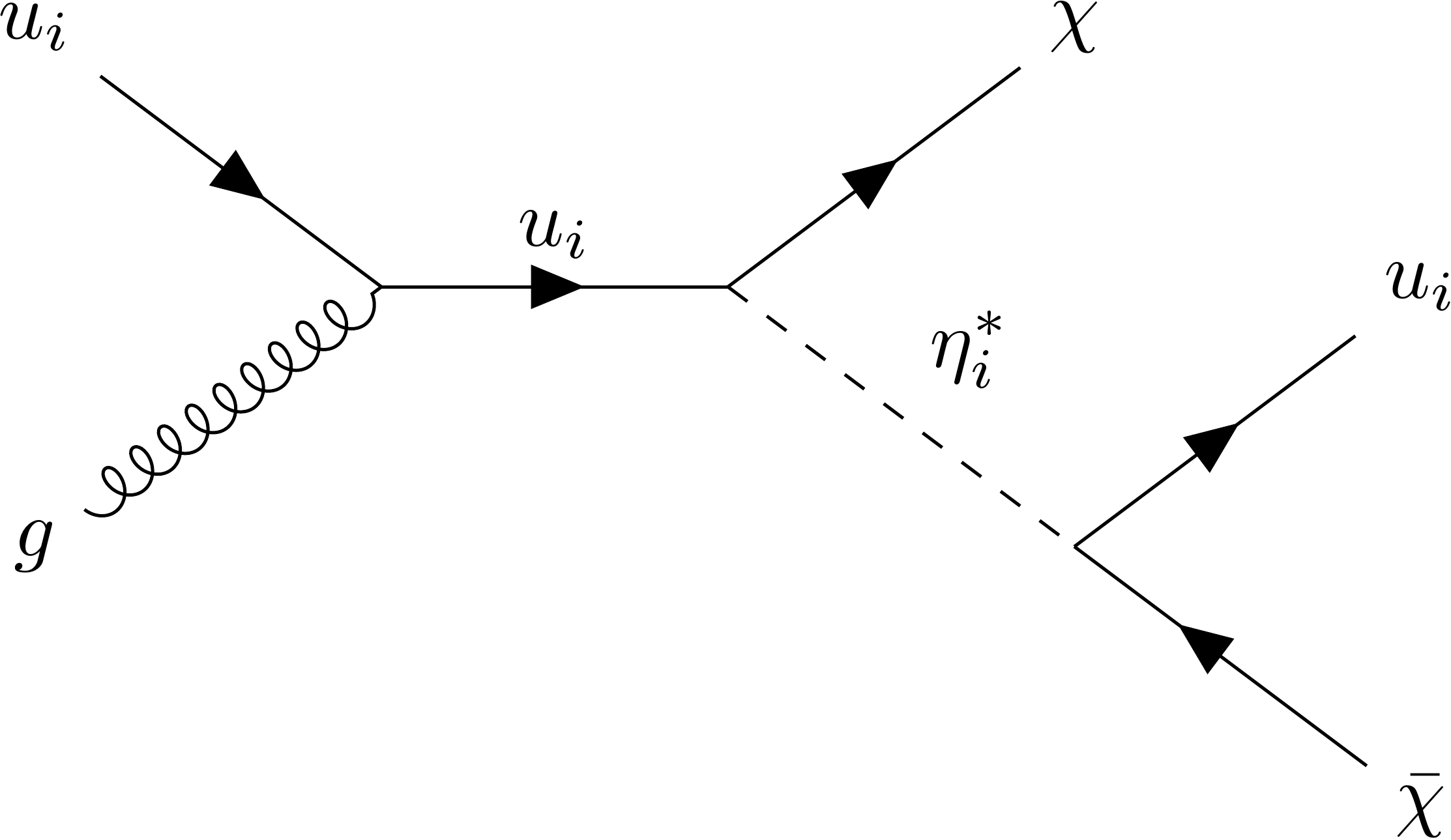}\qquad
\includegraphics[width=0.3\textwidth]{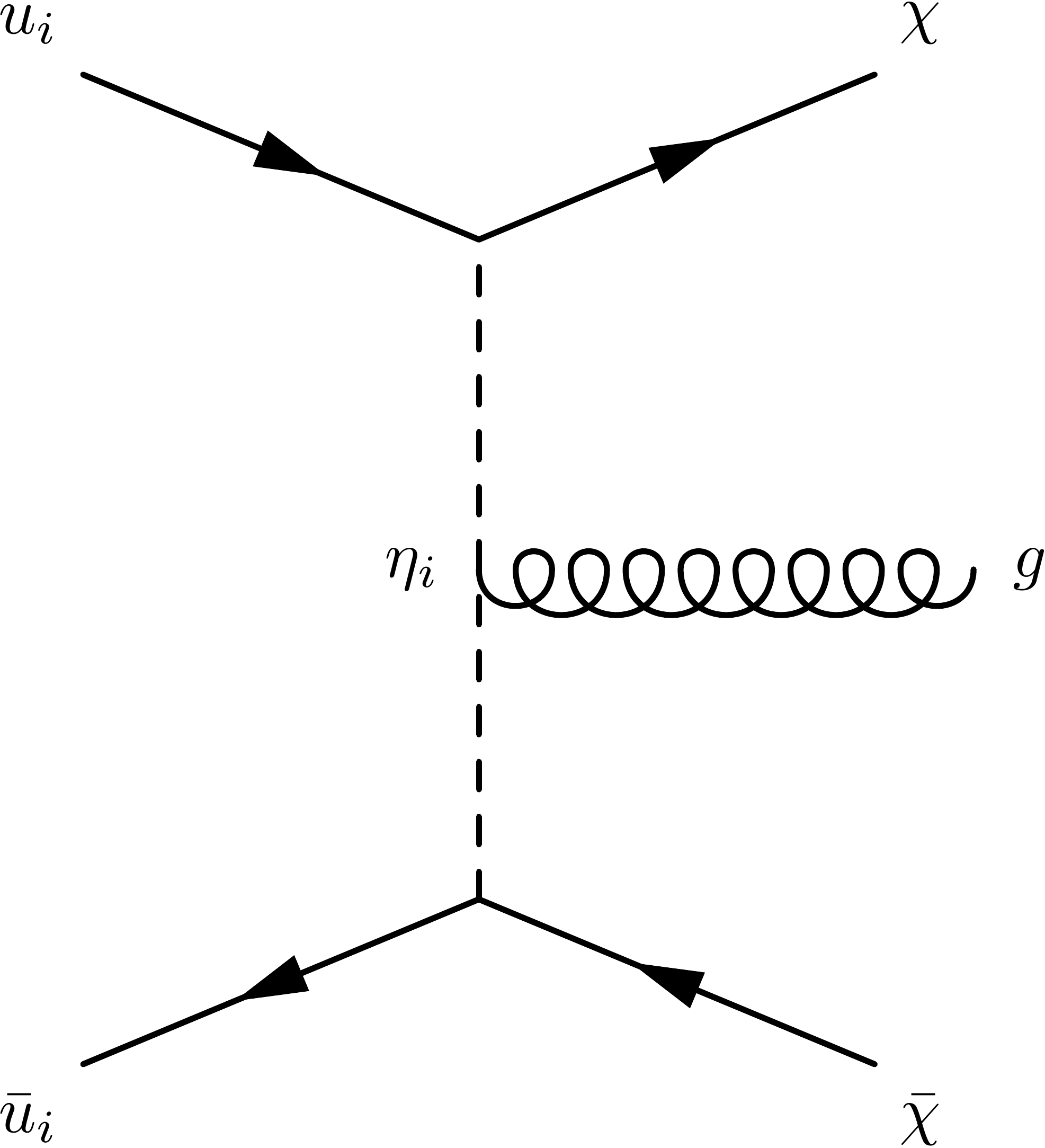}
\caption{\em 
Diagrams contributing to mono-jet signals in $0t\frac{1}{2}$ model. }
\label{fig:0t12monoj}
\end{figure}

Given the similarity of the mediator to squarks,  collider searches for this class of model can fruitfully combine usual mono-jet with strategies for squark detection. 
The main contributions to the $\missET+j$ process come from the diagrams in Fig.~\ref{fig:0t12monoj}, relative
to the processes $u\bar u\to \bar\chi\chi+g$, $ug\to \bar\chi\chi +u$, $\bar u g\to \bar\chi\chi+\bar u$.

Typically, the diagram on the right of Fig.~\ref{fig:0t12monoj} tends to dominate because of larger parton luminosity of the gluon.
The gluon radiation from the $t$-channel mediator is also possible (last diagram of Fig.~\ref{fig:0t12monoj}), but it is suppressed by a further $1/M^2$ (it would correspond to a dimension-8 operator in the low-energy EFT).

\begin{figure}[t!]
\centering
\includegraphics[width=0.35\textwidth]{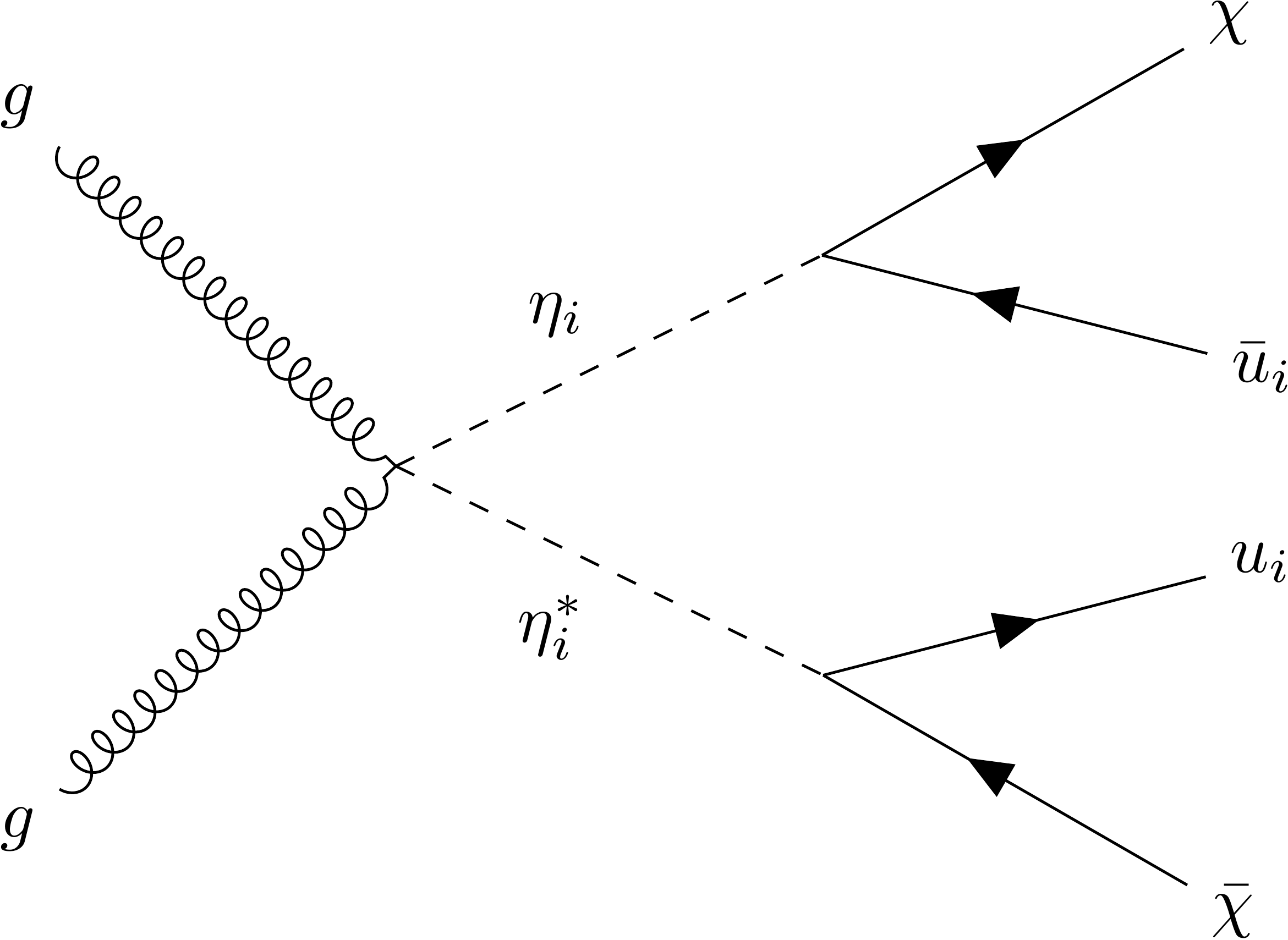}\quad
\includegraphics[width=0.35\textwidth]{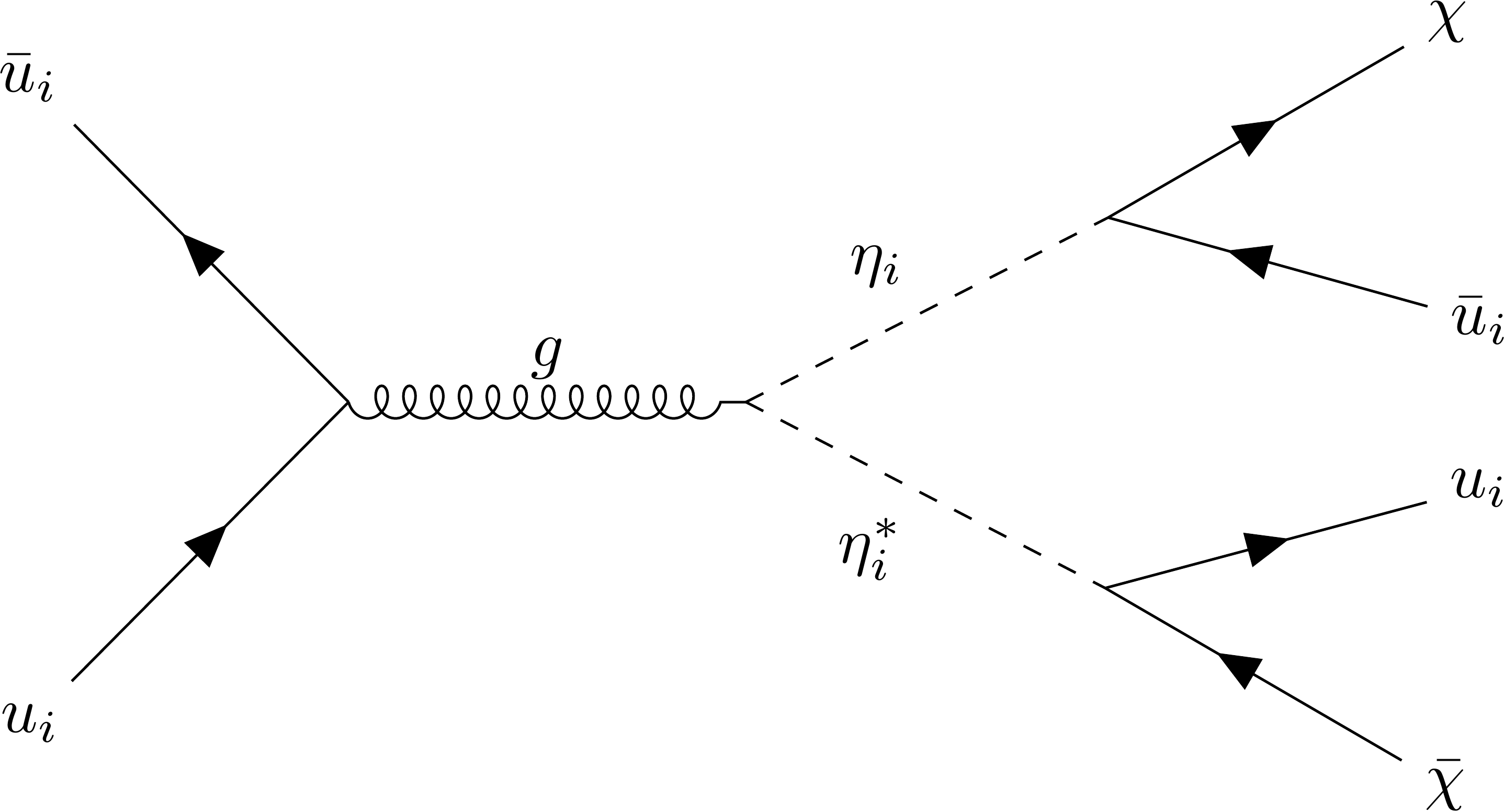}\quad
\includegraphics[width=0.25\textwidth]{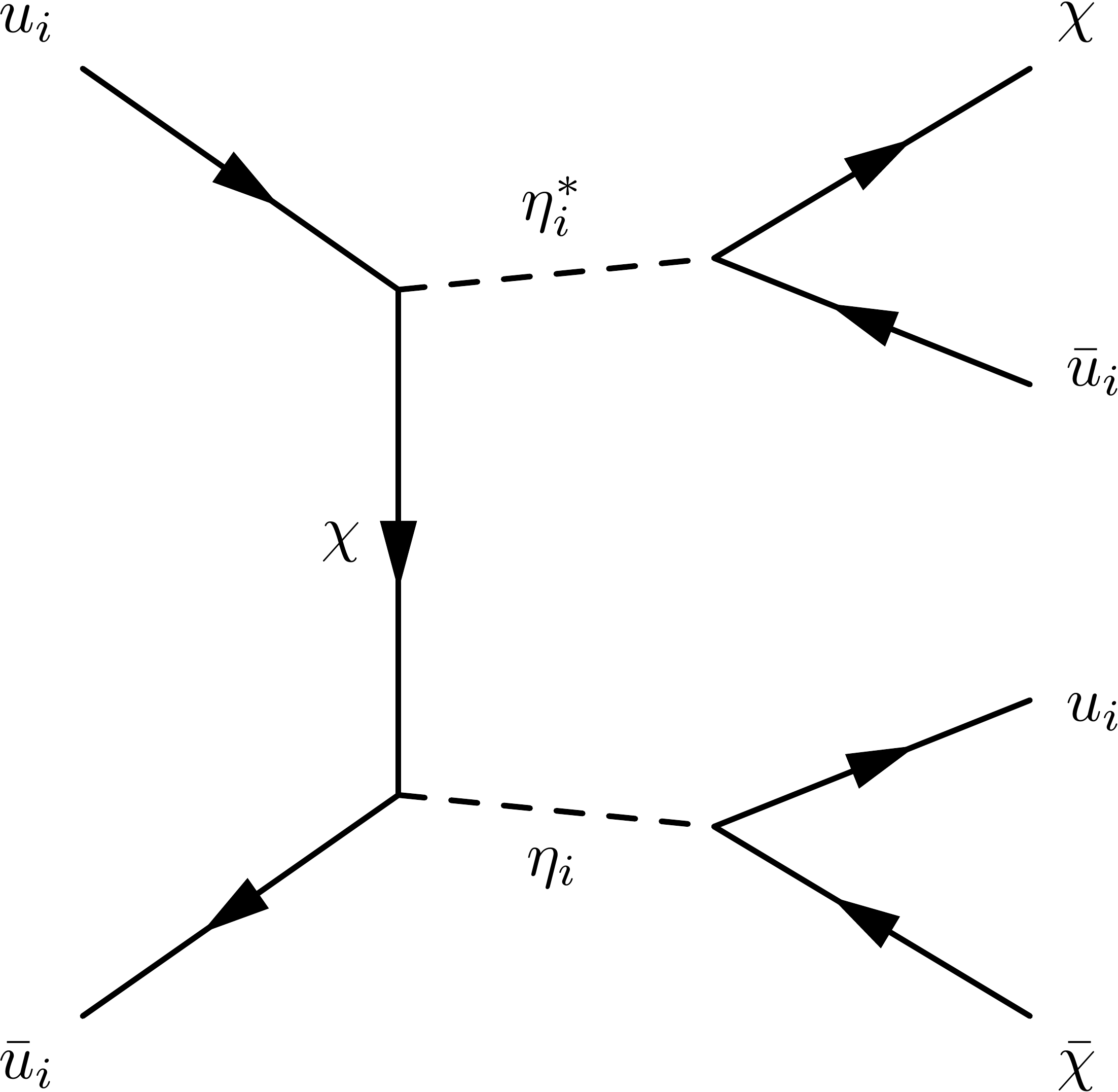}
\caption{\em 
Diagrams contributing to $\missET+2j$ signals in $0t\frac{1}{2}$ model.}
\label{fig:0t12secondj}
\end{figure}

Mono-jet searches allow the possibility of a second jet: $\missET+2j$. These processes are mainly sourced by mediator pair production ($pp\to\eta_1\eta_1^*$) followed by mediator splitting ($\eta_1\to \chi u$), as in Fig.~\ref{fig:0t12secondj}, relative to processes $gg\to \bar\chi \chi \bar u u$, $\bar u u\to \bar\chi \chi \bar u u$.
If the DM is a Majorana particle, further mediator pair production processes are possible, initiated by $uu$ or $\bar u \bar u$ states.

Unlike squark searches, where the squark-neutralino coupling is fixed by supersymmetry to be  weak, in the simplified models $g_1$ is a free parameter. Depending on its magnitude, the relative weights of the diagrams change. For instance, if $g_1$ is weak ($g_1\ll g_s$) the QCD pair production dominates over the production through DM exchange. 

Comprehensive analyses of collider constraints on $t$-channel mediator models with fermion DM have been presented in Refs.~\cite{1308.0592, 1308.0612, 1308.2679, 1402.2285}. The combination of mono-jet and squark searches leads to complementary limits.
The mono-jet searches are usually stronger in the case where the DM and the mediator are very close in mass. 

Before closing this part, it is useful to quote here the result for the mediator width, in the model of Eq.~(\ref{0t12lagr})
\be
\Gamma(\eta_i\to \chi\bar u_i)=\frac{g_i^2}{16\pi}\frac{M_i^2-m_\chi^2-m_{u_i}^2}{M_i^3}
\sqrt{(M_i^2-m_\chi^2-m_{u_i}^2)^2-4m_\chi^2m_{u_i}^2}.
\ee

\paragraph{\textbf{\emph{DM self-annihilation\\}}}

The main process for DM self-annihilations is $\bar\chi\chi\to\bar u_i u_i$, via $t$-channel exchange of the mediator $\eta_i$. This is the relevant process for indirect DM searches.

However, the situation is different for freeze-out calculations. If the DM and the mediator are sufficiently close in mass ($M_i-m_\chi\lesssim T_{\rm freeze-out}$), coannihilations become relevant and one should also take into account  the mediator self-annihilations and the $\chi\eta$ scatterings.
The details of these processes are strongly dependent on whether the DM is a Dirac or Majorana fermion.

For Dirac $\chi$ ($0t\frac{1}{2}_D$ model)
\bea
\langle\sigma v_{\rm rel}\rangle(\bar\chi\chi\to \bar u_i u_i)&=&
\frac{3 g_i^4}{32\pi}
\frac{m_\chi^2}{(m_\chi^2+ M_i^2)^2}\qquad (m_{u_i}=0)\\
\langle\sigma v_{\rm rel}\rangle(\chi\eta_i^*\to u_i g)&=&
\frac{g_s^2 g_i^2}{24\pi}
\frac{1}{M_i(m_\chi+ M_i)}\qquad (m_{u_i}=0)\\
\langle\sigma v_{\rm rel}\rangle(\eta_i\eta_i^*\to g g)&=&
\frac{7 g_s^4}{216\pi}
\frac{1}{M_i^2}\,,
\eea
while the process $\eta_i\eta_i^*\to \bar u_i u_i$ is $p$-wave suppressed.

For Majorana $\chi$ ($0t\frac{1}{2}_M$ model)
\bea
\langle\sigma v_{\rm rel}\rangle(\chi\chi\to \bar u_i u_i)&=&
\frac{g_i^4}{64\pi}
\frac{m_\chi^2 (m_\chi^4+M_i^4)}{(m_\chi^2+ M_i^2)^4}v_{\rm rel}^2\qquad (m_{u_i}=0)
\eea
is $p$-wave suppressed, and
\bea
\langle\sigma v_{\rm rel}\rangle(\chi\eta_i^*\to u_i g)&=&
\frac{g_s^2 g_i^2}{24\pi}
\frac{1}{M_i(m_\chi+ M_i)}\qquad (m_{u_i}=0)\\
\langle\sigma v_{\rm rel}\rangle(\eta_i\eta_i^*\to \bar u_i u_i)&=&
\frac{g_i^4}{6\pi}
\frac{m_\chi^2}{(m_\chi^2+ M_i^2)^2}\qquad (m_{u_i}=0)\\
\langle\sigma v_{\rm rel}\rangle(\eta_i\eta_i^*\to g g)&=&
\frac{7 g_s^4}{216\pi}
\frac{1}{M_i^2}\,.
\eea
The $p$-wave suppressed self-annihilation cross section for Majorana DM bas been thought to be an issue for studying this model with indirect detection. However, it has been noted that the radiation of an EW gauge boson is able to lift the suppression and open up phenomenologically interesting channels for indirect detection \cite{1009.2584,1101.3357,1104.3823,1104.2996}. This has interesting  implications, as the decay of a radiated massive gauge bosons into hadronic final states means that even if the mediator only couples the DM to leptons, photons and antiprotons will inevitably be produced. Electroweak radiation is in general important to take into account when attempting to explain an observed signal such as the apparent excess in the positron flux \cite{Accardo:2014lma,1109.0521} without overproducing other standard model particles such as antiprotons \cite{Aguilar:2015ooa}. 
 This is especially important in the $0t\frac{1}{2}_M$ model when the DM and mediator are near degenerate in mass, as the $2\rightarrow 3$ process $\bar \chi \chi \rightarrow \bar f f' V$ can even dominate over the $2\rightarrow2$ process $\bar \chi \chi \rightarrow \bar f f$.

\paragraph{\textbf{\emph{DM scattering on nucleons\\}}}

As before, the phenomenology is quite different for Dirac and Majorana DM. The DM-nucleon scattering in the low-energy is driven by the effective operator $(\bar\chi u_i)(\bar u_i\chi)$, which can be expanded using Fierz identities into a sum of $s$-channel operators in the chiral basis \cite{1104.3823}
\be
(\bar\chi u_i)(\bar u_i\chi) = \frac{1}{2}(\bar\chi  \gamma^\mu P_L \chi)  (\bar u_i \gamma_\mu P_R u_i)  \sim \Op{D5} - \Op{D6} + \Op{D7} - \Op{D8},
\ee
where $P_L = \frac{1-\gamma_5}{2}$ and $P_R = \frac{1+\gamma_5}{2}$ are the usual chiral projection operators. If $\chi$ is a Dirac fermion, the D5 operator is non-vanishing and provides the spin-independent contribution to the DM-nucleon cross section
\be
\sigma_{\chi N}^{\rm SI}=\frac{g_1^4}{64\pi}\frac{\mu_{\chi N}^2}{(M_1^2-m_\chi^2)^2}f_N^2\qquad (N=n,p)\,,
\ee
where $f_n=1, f_p=2$ because in the Lagrangian Eq.~(\ref{0t12lagr}), $\chi$ scatters only with up-quarks.

If $\chi$ is a Majorana fermion, the D5 and D7 operators vanish identically and the others only contribute to the spin-suppressed scattering operators \Op[NR]{4}, \Op[NR]{8} and \Op[NR]{9}, listed in Table~\ref{NRList}.
For  Dirac DM ($0t\frac12_D$),  limits from the LHC and direct detection turn out to be incompatible with full relic density abundance from thermal freeze-out. On the other hand, the $0t\frac12_M$ model with $m_\chi\gtrsim 100$ GeV is still viable. Of course one should keep in mind that bounds from the relic density are not robust, as the DM may not be thermally produced, or thermal production may make only a fraction of the present DM density.

\subsection{Fermion Mediator}

When the mediator is a fermion, the $2\to 2$ scattering process of a pair of colorless DM particles with two SM particles occurs in the $t$-channel. The DM can either be a scalar ($\frac12t0$ model) or a fermion ($\frac12t\frac12$ model).

\subsubsection{Scalar DM, $t$-channel ($\frac{1}{2}t0$ model)}
\label{subsec:12t0}

If the DM is a SM-singlet scalar $\phi$, it is possible for the mediator to be a vector-like fermion $\psi$ exchanged in the $t$-channel.
Following Ref.~\cite{1511.04452}, we will consider the Lagrangian
\be
\mathcal{L}_{\frac12 t 0}\supset
\frac12 (\partial_\mu\phi)^2-\frac12 m_\phi\phi^2
+\bar\psi(i\slashed{D}-M_\psi)\psi
+(y\phi\bar\psi q_R+\textrm{ h.c.}) \,.
\label{lagr12t0}
\ee
One can choose to couple the DM and the mediator to any SM right-handed or left-handed fermion. The choice made in Eq.~(\ref{lagr12t0}) consists of focusing on couplings to  right-handed quarks, which plays the major role for LHC and direct detection phenomenology (see Refs. \cite{1307.6181, 1307.6480} for the lepton case). The discussion for the case of couplings to $q_L$ would be straightforward. This model has also been mentioned in Ref.~\cite{1308.0592}.

Of course, a singlet scalar DM can also have interactions with the Higgs boson, of the kind discussed in Sect.~\ref{subsec:0s0}. However, in the spirit of the simplified model one usually ignores such interactions when studying the model described by Eq.~(\ref{lagr12t0}). 

By putting together the limits from the LHC, direct detection, indirect detection, thermal relic abundance, and perturbativity of the coupling constant $y$, one finds that this model is rather constrained, but still some parameter space is available, for $m_\phi\gtrsim 1$ TeV and $m_\psi/m_\phi\lesssim 2$ (see Refs.~\cite{1405.6917,1405.6921,1511.04452} for more details).

\paragraph{\textbf{\emph{Collider\\}}}

At the LHC, it is possible to produce a pair of DM particles starting from two quarks with the mediator exchanged in the
$t$-channel, and associated initial-state radiation. This would give the usual mono-jet ($\missET$+j) signal. In addition, if the mediator is light enough, a pair of mediators can be produced, with each of them subsequently decaying into DM and a quark, thus producing an $\missET$   signal in association with 2 or more jets.
One can therefore combine these two kinds of strategies to improve the discovery potential.

Notice that, since the mediator carries color and EW charges, the mediator pair-production can proceed either by DM exchange or by direct QCD and EW Drell-Yan production (see Refs.~\cite{1311.7667, 1505.04306} for experimental results on vector-like quark searches).

For mediator masses $M_\psi$ of the same order as $m_\phi$, the current LHC constraints imply $m_\phi\gtrsim 1$ TeV, but the bounds gets weaker as the mediator mass gets higher \cite{1511.04452}.

\paragraph{\textbf{\emph{DM self-annihilation\\}}}

The main tree-level process for DM self-annihilations is $\phi\phi\to\bar q q$, via $t$-channel exchange of the mediator $\psi$. This is the relevant process to be considered for indirect DM searches.
The thermally-averaged self-annihilation cross section reads \cite{1307.6480, 1511.04452}
\be
\langle\sigma v_{\rm rel}\rangle(\phi\phi\to \bar q q)=
\frac{3y^4}{4\pi}\frac{1}{m_\phi^2\left(1+r^2\right)^2}
\left[
\frac{m_q^2}{m_\phi^2}\left(
1-\frac{2}{3} \frac{1+2r^2}{(1+r^2)^2}v^2
\right)
+\frac{v^4}{15(1+r^2)^2}
\right]
\ee
with $r\equiv {m_\psi}/{m_\phi}>1$. 
Notice the $d$-wave suppression $v^4$, in the case of massless final state particles $m_q=0$, peculiar to real scalar annihilations, and in contrast with the well-known $p$-wave suppression at work when the annihilating particles are Majorana fermions.

The processes of Virtual Internal Bremsstrahlung (radiation of a gluon from the $t$-channel mediator line), or the loop-induced annihilation of $\phi\phi\to gg$ are able to lift the velocity suppression and open up potentially sizeable contributions to the annihilation cross sections. In particular, the one-loop process contributes as $\sigma v\sim r^{-4}$ (but without $m_q$ suppression) while the internal Bremsstrahlung contributes as $\sigma v\sim r^{-8}$.

So, for mediator masses sufficiently close to the DM particle ($r$ close to 1) these higher-order contributions are able to overcome the tree-level process and dominate the annihilation cross section.
However, when the mediator and DM mass are very close, it is also necessary to take into account  the effects of co-annihilations (e.g. $\bar\psi \psi\to \bar q q$) in the early universe. 

As for the general treatment of the annihilations of two particles carrying color, the non-perturbative Sommerfeld effects may play an important role, see Refs.~\cite{1402.6287, 1503.01500, 1511.04452}.

\paragraph{\textbf{\emph{DM scattering on nucleons\\}}}

In this model, the DM scattering on nucleons can proceed by tree-level fundamental interactions of DM with quarks (via exchange of $\psi$), or by loop-induced interactions of DM with gluons. In the former case, integrating out the heavy mediator $\psi$ leads to effective interactions proportional to the quark mass operator and a twist-2 operator
\bea
\mathcal{L}_{\rm eff, 1}&\propto& m_q \phi^2 \bar q q,\\
\mathcal{L}_{\rm eff, 2}&\propto&
 \frac{i}{2}(\partial_\mu\phi)(\partial_\nu \phi)\left[
 \bar q\gamma^\mu \partial^\nu  q
 + \bar q\gamma^\nu \partial^\mu  q
 -(1/2)g^{\mu\nu}\bar q\slashed{\partial}q \,,
 \right]
\eea
while in the latter case, 
\be
\mathcal{L}_{\rm eff, 3}\propto \frac{\alpha_s}{\pi}\phi^2 
\textrm{Tr}[G_{\mu\nu}G^{\mu\nu}]\,.
\ee
The corresponding spin-independent DM-nucleon scattering cross section can be found using Eq.~(\ref{sigSI}), with coefficient \cite{1511.04452, 1502.02244}
\be
c^N=\frac{y^2}{m_\phi^2}\left[
\frac{2r^2-1}{4(r^2-1)^2} f_q^{(N)}
+\frac{3}{4}(q_2^{(N)}+\bar q_2^{(N)})
-\frac{8}{9}\frac{y^2}{24(r^2-1)}f_g^{(N)}
\right]
\ee
where $q_2^{(N)}, \bar q_2^{(N)}$ are the second moments of the PDFs of the parton $q$ in the nucleon $N$, the first term comes from $\mathcal{L}_{\rm eff, 1}$, the second term from $\mathcal{L}_{\rm eff, 2}$ and the last term from the perturbative short-distance  contribution from $\mathcal{L}_{\rm eff, 3}$, where loop momenta are of the order of the DM mass.

\subsubsection{Fermion DM, $t$-channel ($\frac12t\frac{1}{2}$ model)}
\label{subsec:12t12}

\begin{figure}[t!]
\centering
\includegraphics[width=0.3\textwidth]{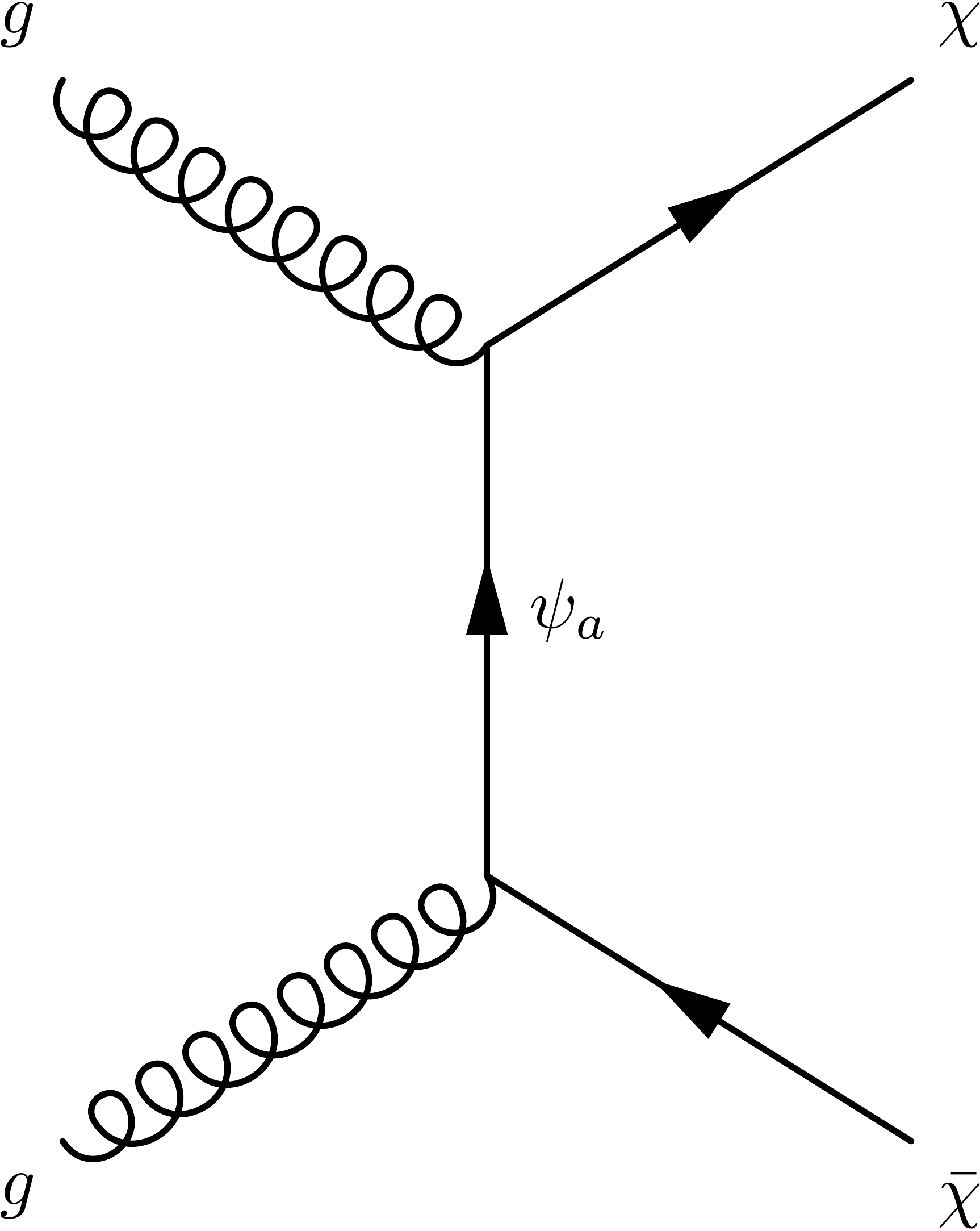}
\caption{\em 
Diagram for DM pair production in $\frac12t\frac12$ model.}
\label{fig:12t12}
\end{figure}

In the case of fermionic DM  with a fermion mediator  exchanged in the $t$-channel, the LHC production can be initiated by two gluons (see tree-level diagram in Fig.~\ref{fig:12t12}). The fermion DM cannot be colored, so the mediator needs to be a
fermion octet (gluino-like) particle $\psi^a$ of mass $M$. 

The operators appearing at the lowest order in the  Lagrangian of the model are
\be
\mathcal{L}_{\frac12 t\frac12}\supset 
\bar\psi^a(i\slashed{D}-M)\psi^a+
\frac{1}{\Lambda}G_{\mu\nu}^a (\bar\psi^a\sigma^{\mu\nu}\chi+\textrm{ h.c.}) 
\label{lagr12t12}
\ee
where $D_\mu$ is the covariant derivative involving the gluon field and the dimension-5 operator is of the form of a chromomagnetic dipole operator (resembling the gluino-gluon-bino interaction in SUSY).

Extensive searches are performed for this kind of mediator, driven by the interest in SUSY models. 
Limits from direct QCD production of gluino-like  mediators decaying to two gluons and two DM particles tell us that the mediator must be heavier than about 1150 GeV (95\% CL) for DM masses below 100 GeV  \cite{1507.05525}.

However, apart from the direct mediator searches,  no analyses have been performed to study the fermion octet in the context of a simplified model with a DM particle, to our knowledge. Of course, the dimension-5 interaction
in Eq.~(\ref{lagr12t12}) would lead to rather weak signals at LHC.
But a careful study of this model, also in view of possible future colliders, would be interesting.

\subsection{Vector Mediator}

With a vector mediator, often labelled $Z'$, it is possible to produce a DM pair from an initial state of two quarks by  exchanging the mediator in the $s$-channel, with DM being a scalar ($1s0$ model) or a fermion ($1s\frac12$ model), or in the $t$-channel, with fermion  DM ($1t\frac12$ model). 

We will consider the vector mediator as having an explicit mass, without trying to justify it from a more complete UV theory, following the 
 philosophy behind simplified models. It is assumed that there exists some UV completion that can avoid problems of gauge invariance, anomaly cancellation and mass generation; and importantly, that the phenomenology is independent of the UV completion. However, care must be taken, since this is not always the case. Some choices of parameters within simplified models can be pathological, such that no fully consistent UV completion exists. This is the case for a fully axial-vector model, where the model violates gauge invariance unless the SM particles also couple to the mediator via a vector coupling \cite{1510.02110}. This can lead to unphysical signals in regions where the model violates perturbative unitarity \cite{1503.05916,1512.00476,1502.05721,1603.01366}.

\subsubsection{Scalar DM, $s$-channel ($1s0$ model)}
\label{subsec:1s0}

For a complex scalar DM $\phi$ of mass $m_\phi$ coupled to the vector mediator $V_\mu$ (often labelled $Z'$) of mass $M_V$, the Lagrangian of the model is given by
\be
\mathcal{L}_{1s0}\supset
-V^\mu\left[
g_\phi\left[\phi^*(i\partial_\mu\phi)-\phi(i\partial_\mu\phi^*)\right]
+
\sum_f
\bar f\gamma_\mu(g_f^V+g_f^A\gamma^5)f
\right]\,,
\label{1s0lagr}
\ee
where the sum over $f$ extends to all SM fermions.

The couplings $g_f^{V,A}$ need to be flavor independent in order to respect MFV hypothesis.  It is customary in the literature to reduce the number of free parameters by considering only the limiting cases of a ``purely vector'' ($g_i^A=0$) or a ``purely axial'' ($g_i^V=0$) mediator.

\paragraph{\textbf{\emph{Collider\\}}}

The collider phenomenology of this class of models is crucially dependent on the leading decay channels of the vector mediator, provided they are kinematically accessible. 
The decay width of $V$ to SM fermions $f$, with color number $N_c(f)$, is given by
\be
\Gamma(V\to \bar f f)=N_c(f)\frac{M_V}{12\pi}\sqrt{1-\frac{4m_f^2}{M_V^2}}
\left[
|g_f^V|^2\left(1+\frac{2m_f^2}{M_V^2}\right)
+|g_f^A|^2\left(1-\frac{4m_f^2}{M_V^2}\right)
\right]\,,
\label{Vtoff}
\ee
while the (invisible) decay width to DM particles is
\be
\Gamma(V\to \phi\phi)=\frac{g_\phi^2 M_V}{48\pi}\left(1-\frac{4m_\phi^2}{m_V^2}\right)^{3/2}\,.
\ee
Roughly speaking, if invisible decays dominate ($V\to\phi\phi$), we expect the collider phenomenology to be driven by MET searches (e.g. mono-jet); conversely, if the mediator predominantly decays to SM fermions, the best search strategy would be the heavy resonances (e.g. di-jets 
\cite{1503.05916} or di-leptons, although the latter case is highly constrained \cite{1401.0221, 1403.4837}).

Further constraints arise from requiring a particle interpretation of the mediator (narrow-width approximation): $\Gamma_V/M_V<1$.

\paragraph{\textbf{\emph{DM self-annihilation\\}}}

The DM self-annihilation cross section, to be used for relic density calculations or for indirect detection, is
\be
\langle\sigma v_{\rm rel}\rangle(\phi\phi\to  f\bar f)=
N_c(f)\frac{g_\phi^2}{6\pi}
\frac{m_\phi^2\sqrt{1-\frac{m_f^2}{m_\phi^2}}}{(4m_\phi^2-M_V^2)^2}
\left[
|g_f^V|^2\left(1+\frac{1}{2}\frac{m_f^2}{m_\phi^2}\right)+
|g_f^A|^2\left(1-\frac{m_f^2}{m_\phi^2}\right)
\right]v_{\rm rel}^2
\,,
\ee
which is in $p$-wave.

\paragraph{\textbf{\emph{DM scattering on nucleons\\}}}

At low-energies, the DM-nucleon scattering is described by the effective
operator
\bea
\mathcal{L}_{\rm eff}&=&\sum_q \frac{g_\phi}{M_V^2}
\left[\phi^*(i\partial_\mu\phi)-\phi(i\partial_\mu\phi^*)\right]
\left[\bar q\gamma^\mu(g_q^V+g_q^A\gamma^5)q\right]\nn\\
&\simeq &\sum_q\frac{g_\phi g_q^V}{M_V^2}
\left[\phi^*(i\partial_\mu\phi)-\phi(i\partial_\mu\phi^*)\right]
\left[\bar q\gamma^\mu q \right]
\eea
where the axial contribution has been neglected. Notice, however, that the operator mixing due to the RGE flow would generate a vector contribution even starting from a purely axial term \cite{1605.04917}.
 The spin-independent component of the cross section can be found using Eq.~(\ref{sigSI}), with coefficients (cf. Table \ref{ScalarDMNucleonList})
\be
c^p=\frac{1}{M_V^2}g_\phi(2g_u^V+g_d^V) \,,
\qquad
c^n=\frac{1}{M_V^2}g_\phi(2g_d^V+g_u^V)\,.
\ee

\subsubsection{Fermion DM, $s$-channel ($1s\frac{1}{2}$ model)}
\label{subsec:1s12}

This class of models has been studied extensively; For a non-exhaustive list, see Refs.~\cite{hep-ph/0408098,0803.4005,0809.2849,1106.0885,1107.2118,1112.5457,1202.2894,1204.3839,1212.2221,1312.5281,1401.0221,1403.4837,1406.3288,1407.8257,1411.0535,1501.03490,1502.05721,1503.05916,1505.05710,1506.06767,1509.07867,1510.02110,1512.00476,1605.06513,1605.07940}. 
The Lagrangian of the model is given by
\be
\mathcal{L}_{1s\frac{1}{2}}\supset
-V^\mu\left[
\bar\chi\gamma_\mu(g_\chi^V+g_\chi^A\gamma^5)\chi
+
\sum_f
\bar f\gamma_\mu(g_f^V+g_f^A\gamma^5)f
\right]\,,
\label{1s12lagr}
\ee
where the sum over $f$ extends to all SM fermions.
If $\chi$ is Majorana, the vector bilinears vanish identically, so $g_\chi^V=0$.

The MFV hypothesis imposes the couplings $g_f^{V,A}$ to be flavor independent.  In the most general case, there are several model parameters, therefore  a ``purely vector'' ($g_i^A=0$) or a ``purely axial'' ($g_i^V=0$) mediator is often assumed in the literature.

\paragraph{\textbf{\emph{Collider\\}}}

The collider phenomenology of the mediator is the same as the one already discussed for the $1s0$ model, except that the invisible width of the mediator
is now given by the same expression as the decay to SM fermions Eq.~(\ref{Vtoff}) with the index $f$ replaced by the index  $\chi$ and $N_c(\chi)=1$.

\paragraph{\textbf{\emph{DM self-annihilation\\}}}

The dominant ($s$-wave) contribution to the DM self annihilation cross-section is 
\bea
\langle\sigma v_{\rm rel}\rangle(\chi\chi\to  f\bar f) &=&
\frac{N_c(f)}{2\pi}
\frac{m_\chi^2\sqrt{1-\frac{m_f^2}{m_\chi^2}}}{(4m_\chi^2-M_V^2)^2}
\Bigg\{
|g_\chi^V|^2\left[
|g_f^V|^2\left(2+\frac{m_f^2}{m_\chi^2}\right)+
2|g_f^A|^2\left(1-\frac{m_f^2}{m_\chi^2}\right)\right]\nn\\
&&
+|g_\chi^A|^2|g_f^A|^2\frac{m_f^2}{m_\chi^2}\left(1-\frac{4m_\chi^2}{m_V^2}\right)^2
\Bigg\},
\eea
where the term proportional to $|g_\chi^A|^2|g_f^V|^2$ is absent here because it appears only at the level of $p$-wave.

\paragraph{\textbf{\emph{DM scattering on nucleons\\}}}

In the low-energy limit, the effective interactions relevant for DM-nucleon scatterings are
\be
\mathcal{L}_{\rm eff}=\sum_q \frac{1}{M_V^2}
\left[\bar\chi\gamma_\mu(g_\chi^V+g_\chi^A\gamma^5)\chi\right]
\left[\bar q\gamma^\mu(g_q^V+g_q^A\gamma^5)q\right]\\
\ee
The $g_\chi^V g_q^V$ terms lead to a SI cross section, while the purely axial terms proportional to $g_\chi^A g_q^A$ lead to SD scattering. The cross terms $g_\chi^V g_q^A, g_\chi^A g_q^V$ give cross sections suppressed by either the DM velocity or the momentum, so they are subdominant and can be neglected.
Again, it should be noted that, because of operator mixing induced by the RGE flow, the axial and vector quark currents mix and the term proportional to $g_\chi^V g_q^A $ would also contribute to the dominant term $g_\chi^V g_q^V$.
The spin-independent component of the cross-section is given by Eq.~(\ref{sigSI}) with coefficients (cf. Table \ref{DMNucleonList})
\be
c^p=\frac{1}{M_V^2}g_\chi^V(2g_u^V+g_d^V) \,,
\qquad
c^n=\frac{1}{M_V^2}g_\chi^V(2g_d^V+g_u^V)\,,
\ee
and the spin-dependent component by Eq.~(\ref{sigSD}) with coefficient 
(cf. Table \ref{DMNucleonList})
\be
c^N=\frac{1}{M_V^2}\sum_{q=u,d,s} g_\chi^A g_q^A
\Delta_q^{(N)}\,,
\ee
where  sample values for $\Delta_q^{(N)}$ are given in Table~\ref{DDConstants}.

\paragraph{\textsf{\emph{$\blacksquare$ CASE STUDY 3: $Z$ AS MEDIATOR\\}}}

The SM $Z$ boson itself may serve as a vector mediator, rather than a specultative particle. In this case, the couplings $g_f^{V,A}$ of the $Z$ boson to SM fermions are well-known: $g_V=(g_2/\cos\theta_W)(1/4-(2/3)\sin^2\theta_W)$, $g_A=-g_2/(4\cos\theta_W)$ for up-type quarks, and
$g_V=(g_2/\cos\theta_W)(-1/4+(1/3)\sin^2\theta_W)$, $g_A=g_2/(4\cos\theta_W)$ for down-type quarks, 
 where $g_2$ is the $SU(2)_L$ gauge coupling and $\theta_W$ is the weak mixing angle.
 
The Lagrangian has the same form as that of a generic vector mediator Eq.~(\ref{1s0lagr}) for scalar DM or Eq.~(\ref{1s12lagr}) for fermion DM, therefore all the results listed in Sections \ref{subsec:1s0} and  
\ref{subsec:1s12} apply, 
except that the $Z$ couplings to fermions $g_f^{V,A}$ are known.

Let us summarize the main points of the analysis carried out in Ref.~\cite{1402.6287}, to which we refer the reader for further details. In the mass regime where $Z$-decays to DM are kinematically allowed ($m_\chi<M_Z/2$), 
the experimental constraint on the $Z$ invisible width $\Gamma_{Z,inv}\lesssim 2$ MeV gives 
$g_\phi\lesssim 0.08 (g_2/\cos\theta_W)$ and 
$g_\chi^{V,A}\lesssim 0.04 (g_2/\cos\theta_W)$.

The opposite mass regime $m_\chi>M_Z/2$ is not significantly constrained by collider data with respect to the much stronger constraints coming from direct detection.

Indeed, direct detection experiments (currently dominated by LUX results), place quite strong  limits on $g_\chi^V, g_\phi\lesssim 10^{-3}(g_2/\cos\theta_W)$ for DM masses around 100 GeV, while the spin-dependent interactions lead to a milder bound on $g_\chi^A\lesssim 0.3 (g_2/\cos\theta_W)$ for DM mass around 100 GeV.

As far as the thermal relic density is concerned, a scalar thermal DM candidate accounting for 100\% of the DM abundance is ruled out, for $m_\phi\lesssim$ TeV. As for fermion DM, the pure vector case ($g_\chi^A=0$)
is still compatible with direct detection and relic abundnace for DM masses above about 1 TeV (and near the resonance region $m_\chi\simeq M_Z/2$), while a thermal DM candidate with  pure axial couplings to the $Z$ ($g_\chi^V=0$) is still viable in most of the parameter space with $m_\chi>M_Z/2$.

However, It should  be kept in mind that the conclusions drawn above are only valid  within the simple model described by the SM plus the DM particle;  new physics particles and interactions at the weak scale can have a big impact on the bounds from relic density.

\paragraph{\textsf{\emph{$\blacksquare$ CASE STUDY 4: A SUSY-INSPIRED EXAMPLE, SINGLET-DOUBLET DM\\}}}
\label{casestudy4}
A different  possibility is to allow mixing between an EW singlet and an EW doublet as a mechanism to generate interactions between the dark and the visible sectors \cite{hep-ph/0510064, 0705.4493, 0706.0918, 1109.2604,1505.03867} (see also Refs.~\cite{1403.7744,1601.01354} for alternative electroweak representations).  Such a situation is  also interesting  because it can  be realized in SUSY with a bino-higgsino mixing, in the decoupling limit where the masses of the scalar superpartners and of the wino are much larger than $M_1$ and $|\mu|$.

The particle content of the model consists of a fermion singlet $\chi$ and two fermion doublets $\Psi_1=(\Psi_1^0, \Psi_1^-)^T$ and $\Psi_2=(\Psi_2^+, \Psi_2^0)^T$, with opposite hypercharges. There is a discrete $Z_2$ symmetry under which $\chi, \Psi_1, \Psi_2$ are odd while the SM particles are even. The Lagrangian describing the interactions is given by
\be
\mathcal{L}=\bar\chi(i\slashed\partial)\chi+\sum_{i=1,2}\bar\Psi_i(i\slashed D)\Psi_i-\frac12 (\chi, \Psi_1, \Psi_2)\mathcal{M}(\chi, \Psi_1, \Psi_2)^T
\ee
where $D_\mu$ is the covariant derivative and the mass matrix is
\be
\mathcal{M}=
\left(\begin{matrix}
m_S & y_1\frac{v}{\sqrt{2}} & y_2\frac{v}{\sqrt{2}}\\
y_1\frac{v}{\sqrt{2}} & 0 & m_D\\
y_2\frac{v}{\sqrt{2}} & m_D & 0
\end{matrix}\right)
\ee
with $m_S, m_D$ the mass parameters for the singlet and doublet, respectively. The off-diagonal singlet-doublet mixing terms arise from interaction terms with the Higgs (after EW symmetry breaking) of the kind $ -\chi  (y_1 H \Psi_1+y_2 H^\dag \Psi_2)+ \textrm{ h.c.}$.

The diagonalization of the mass matrix via the unitary matrix $U$ performs the shift to the mass-eigenstates basis where the physical spectrum of the model becomes apparent: 2 charged states $\chi^\pm$ and 3 neutral states $\chi_{1,2,3}$ such as
\be
\left(\begin{matrix}
\chi_1\\
\chi_2\\
\chi_3
\end{matrix}\right)
=U
\left(\begin{matrix}
\chi\\
\Psi_1\\
\Psi_2
\end{matrix}\right)
\ee
with the lightest neutral state $\chi$ playing the role of the DM particle.

In the  language of SUSY, the lightest neutralino coming from the mixing with bino-higgsino states is the DM.
One can recover the SUSY situation with the following identifications: $m_S=M_1, m_D=|\mu|$, $y_1=-\cos\beta g_1/\sqrt{2}$
and $y_2=\sin\beta g_1/\sqrt{2}$, where $g_1$ is the $U(1)_Y$ gauge coupling and $\beta$ is the misalignment angle between the VEVs of $H_u$ and $H_d$: $\tan\beta=v_u/v_d$.

In the mass-eigenstates basis it is also easy to read the interactions between the new states and the SM bosons (physical Higgs $h$ and $Z,W^\pm$)
\bea
\mathcal{L}&\supset& 
-h\bar\chi_i(\Re(c^h_{ij})+\Im(c^h_{ij})\gamma^5)\chi_j
-Z_\mu\bar\chi_i\gamma^\mu(\Im(c^Z_{ij})-\Re(c^Z_{ij})\gamma^5)\chi_j
\nn\\
&&-\frac{g_2}{2\sqrt{2}}\left(
(U_{i3}-U_{i2}^*)W^-_\mu\bar\chi_i\gamma^\mu \chi^+
-(U_{i3}+U_{i2}^*)W^-_\mu\bar\chi_i\gamma^\mu 
\gamma^5\chi^+  +\textrm{ h.c.}
\right)
\label{singletdoubletlagr}
\,,
\eea
with $i,j=1,3$ and where the couplings to $h$ and $Z$ are
\be
c^h_{ij}=\frac{1}{\sqrt{2}}(y_1 U_{i2}U_{j1}+y_2 U_{i3}U_{j1})\,,
\qquad
c^Z_{ij}=\frac{g_2}{4\cos\theta_W}(U_{i3}U_{j3}^*-U_{i2}U_{j2}^*)
\qquad
\ee
Notice that the DM coupling to $Z$ boson $c^Z_{11}$ has no imaginary part, leading to  a purely axial-vector interaction, and therefore to a spin-suppressed cross section of DM with nucleons, arising from a mix of operators \Op[NR]{4}, \Op[NR]{8} and \Op[NR]{9}.

As we see, this model generates a somewhat hybrid situation given by a combination of 
$0s\frac12$ and $1s\frac12$ models, where the mediation from the dark to the visible sector is provided by both the Higgs and the $W,Z$ bosons.

The self-annihilations of DM proceed via $s$-channel exchange of a Higgs or a $Z$ boson, to a fermion-antifermion final state. But it is also possible for DM to exchange a $\chi_i$ or a $\chi^\pm$ in the $t$-channel to lead to $hh, ZZ, WW$ final states.

If kinematically open, the interactions in Eq.~(\ref{singletdoubletlagr}) contribute to the invisible width of $h$ and $Z$, as 
\bea
\Gamma(h\to \chi_1\chi_1)&=&\frac{|c^h_{11}|^2}{4\pi}m_h\left(
1-\frac{4m_{\chi_1}^2}{m_h^2}
\right)^{3/2}\\
\Gamma(Z\to \chi_1\chi_1)&=&\frac{|c^Z_{11}|^2}{6\pi}m_Z\left(
1-\frac{4m_{\chi_1}^2}{m_Z^2}
\right)^{3/2}
\eea
and the limits on these widths can be used to place bounds on the parameter space

At the LHC, there is a richer phenomenology due to the presence of more (also charged) states. Indeed, in addition to a top-loop-induced gluon fusion process $gg\to\chi_i\chi_j$ there is also a Drell-Yan-type production via EW bosons which opens production modes of the kind $q\bar q\to \chi_i\chi_j, \chi^+\chi^-$ ($Z$-exchange) or $q\bar q\to \chi_i\chi^\pm$ ($W$-exchange).
The further decay of the heavier part of the spectrum $\chi^\pm, \chi_{2,3}$ to the lightest DM particle $\chi_1$ involves further gauge boson radiation with the possibility of lepton-rich final states (such as $2\ell+\missET$ or $3\ell+\missET$), offering clean handles for searches.

\subsubsection{Fermion DM, $t$-channel ($1t\frac{1}{2}$ model)}
\label{subsec:1t12}

At tree-level, it is possible to produce a pair of fermion-DM particles by two initial-state quarks exchanging a vector mediator in the $t$-channel.
In order to preserve the color-, flavor- and charge-neutrality of DM, the 
mediator should carry flavor, color and electric charge. In particular, it must be a color-triplet.

The corresponding LHC phenomenology has some similarities with that of the $0t\frac12$ model (squark-like mediator), as  similar diagrams contribute to the mono-jet signal. But on the other hand, the direct production of the mediator would be different, because of its quantum numbers under $SU(3)_c$ and Lorentz.  

As for the $\frac12t\frac12$ model,  to our knowledge there have been no analyses of the phenomenology of a color-triplet mediator  in the context of a simplified model with a DM particle.

%% file: sections/conclusion.tex

\section{Conclusions}
\label{sec:conclusions}

In this review we have discussed and compared  two important frameworks to describe the phenomenology of particle (WIMP) DM  and simultaneously keep the number of parameters as minimal as possible: the EFT approach and simplified models.

Both of these approaches have virtues and drawbacks, but it is now clear that the use of EFTs in collider searches for DM suffers from important limitations.
Therefore, simplified models are a compelling candidate for providing a simple common language to describe the different aspects of DM phenomenology (collider, direct and indirect searches).

Of course, this does not mean that alternative approaches are not possible or not interesting, and by no means this state-of-the-art review should be regarded as  exhaustive. 
The subject is currently rapidly changing and expanding, in response to an ever-increasing interest in the problem of the  identification of DM. 

We have provided an overview of the subject of EFTs for DM searches, spelling out
the theoretical issues involved in its use but also its advantages. For each effective operator, we also highlighted how to make the connection among the different search strategies.

In the Section dedicated to  simplified models, we provided
a general classification of the models, and proposed a simple nomenclature system for them (cf. Table \ref{SimpModelsTable}). Wherever available, we collected the main results regarding the application of the simplified model to describe the phenomenology of DM production at collider, DM self-annihilations and DM scattering with nuclei. We also emphasized, to the best of our knowledge, which models have been least addressed in the literature, encouraging work to fill these gaps.

By interpreting the results of the different DM searches within a single theoretical framework, such as the one provided by  simplified models, it is possible to dramatically increase the discovery potential and make the discovery of DM more accessible.